\documentclass[structabstract]{aa}  

\usepackage{graphicx}
\usepackage{longtable}
\usepackage{txfonts}
\usepackage{natbib}
\usepackage{multirow}
\usepackage{layout}
\usepackage{url}
\usepackage{nicefrac}
\usepackage{keyval}
\usepackage{wasysym}
\usepackage[colorlinks=false,pdfborder={0 0 0}]{hyperref}
\bibpunct{(}{)}{;}{a}{}{,}

\begin{document}

   \title{RX J1548.9+0851, a fossil cluster?}
   \titlerunning {RX J1548.9+0851, a fossil cluster?}

   \author{P. Eigenthaler  \and W.W. Zeilinger}
   \institute{Institut f\"ur Astronomie, Universit\"at Wien, T\"urkenschanzstra\ss e 17, A-1180 Vienna\\ 
    \email{paul.eigenthaler@univie.ac.at}}

   \date{Received xxx; accepted xxx}

   \abstract
   {Fossil galaxy groups  are  spatially   extended  X-ray  sources with   X-ray luminosities above  $L_{X,\,\textrm{\scriptsize bol}} \ge 10^{42}$   h$_{50}^{-2}$   ergs  s$^{-1}$  and a central  elliptical galaxy
   dominating the  optical,  the  second-brightest galaxy   being  at  least    2  magnitudes  fainter   in the  $R$   band.  Whether these   systems are  a  distinct class  of objects  resulting from
   exceptional formation  and evolution  histories is  still unclear,  mainly   due     to  the  small  number  of objects  studied so  far, mostly  lacking  spectroscopy of  group members   for group
   membership confirmation and a detailed kinematical analysis.}
   {To complement the scarce sample  of spectroscopically studied fossils  down  to their faint  galaxy populations,   the fossil candidate   RX~J1548.9+0851 ($z=0.072$) is  studied in this work.  Our
   results  are compared  with existing data from fossils in the literature.}
   {We use ESO  VLT VIMOS multi-object   spectroscopy to determine  redshifts of the   faint galaxy population  and study the   luminosity-weighted dynamics and  luminosity function of the system. The
   full-spectrum fitting package  ULySS is used to determine ages and metallicities of group members. VIMOS imaging data are used to study the morphology of the central elliptical.}
   {We identify  40 group  members spectroscopically  within the  central $\sim$  300 kpc  of the  system and  find  31  additional redshifts    from the  literature, resulting in a total number of 54
   spectroscopically confirmed  group members within 1 Mpc.  RX~J1548.9+0851 is made up of two bright ellipticals in the central region with a magnitude gap of $\Delta    m_{1,2} =  1.34$ in the  SDSS
   $r'$ band leaving the  definition of  RX~J1548.9+0851 being a fossil to the  assumption of the virial radius. We  find a luminosity-weighted velocity dispersion of  568~km~s$^{-1}$ and  a mass   of
   $\sim 2.5 \times 10^{14}$ M$_{\odot}$   for the   system confirming previous studies that revealed fossils to be massive. An average mass-to-light ratio of $M/L\sim 400$ M$_{\odot}/$L$_{\odot}$  is
   derived from the SDSS $g'$, $r'$, and  $i'$ bands. The central elliptical  is  well-fit  by  a pure  deVaucouleurs    $r^{1/4}$ law without a  cD envelope. Symmetric   shells are  revealed  along
   the major axis  of the galaxy     indicating    a   recent   minor merger. RX~J1548.9+0851  shows   a  steep, increasing    luminosity  function  with a faint-end slope of   $\alpha=     -1.4   \pm
   0.1$.  Satellite galaxies show a clear spatial segregation with respect to their stellar populations -- objects with old stars  are  confined   to an  elongated,  central  distribution
   aligned   with  the major  axis   of the central  elliptical.}
   {Although RX~J1548.9+0851   shows similar  properties compared  to  other  fossils studied  recently, it  might not  be a  fossil at  all, being dominated by 2 bright central ellipticals. Comparing
   RX~J1548.9+0851 with scaling relations  from  ordinary poor groups  and clusters confirm the  idea  that fossils might  simply be normal  clusters  with the richness and  optical luminosity of poor
   groups.}


   \maketitle


\section{Introduction}
\citet{jones03} defined fossil galaxy  groups to be  spatially   extended X-ray   sources with  X-ray luminosities above $L_{X,\,\textrm{\scriptsize  bol}}  \ge 10^{42}$   h$_{50}^{-2}$   ergs   s$^{
-1}$ and a  central elliptical  galaxy dominating the  optical, all  other  galaxies within  one-half  virial  radius  being  at least   2   magnitudes fainter in  the $R$   band. These  observational
characteristics suggested    that  fossils are  the  remnants    of    poor   groups    of    galaxies   that    depleted   their   $L^{*}$     galaxies    in  the course  of dynamical   friction  and
merging, building  up one     massive   central  galaxy  as   predicted by  numerical simulations   (\citealt{barnes89}).     The   first  fossil     was identified    in    1994 (\citealt{ponman94}).
Since                 then,              several            other             objects               have            been             assigned               to                 this                class
(\citealt{vikhlinin,jones03,yoshioka,sun04,khosroshahi04,ulmer,schirmer10,pierini11,miller11}).       \citet{santos}       and \citet{eigenthaler09}       have cross-correlated        optical      and
X-ray         data     from     the      Sloan  Digital Sky    Survey SDSS (\citealt{adelman})   and   the  ROSAT    All    Sky Survey   (\citealt{voges})   to systematically   identify new     fossil
candidates.       \citet{jones03}      estimated    that  fossil  systems  probably  constitute          8-20\%     of    all   systems with       comparable    X-ray  luminosity    ($\ge$10$^{42}$  ergs
s$^{-1}$), acting        also    very    likely   as      the  formation   site  of   brightest cluster    galaxies    (BCGs).   Similar optical   luminosities   of     the  central ellipticals     in
fossils   compared    to     that    of   BCGs    support    this     idea (\citealt{khosroshahi06}).   The detection of  these    aggregates is challenging  however, mainly  because  of  their unremarkable appearance in   the
optical  and the   lack of high $S/N$ X-ray data.

Recent   studies  have   shown  that  fossils   are   much  more   massive   than what  would be expected  from the  remnant   of  a group-sized   halo,  exhibiting  velocity   dispersions  and  X-ray
temperatures   comparable   to  poor  clusters    rather    than   groups  (\citealt{mendesdeoliveira06,mendesdeoliveira09,cypriano,proctor11}),  suppressing  strong  interactions   between   galaxies
-- inconsistent with the  proposed merging   scenario. Hence,  there  has  been a  lot  of  debate  on   the origin  and formation   scenario for  fossil  groups/clusters  over the past  few years.  A
few  previous  studies  have  investigated   the  scaling  relations    of  known   fossils   in  comparison   to  ordinary    groups  and  clusters.  \citet{khosroshahi07} concluded  that fossils are
overluminous in X-rays  for  a  given   optical  luminosity,   falling   on  the  $L_{X}-T_{X}$    relation   however.   This excess   in  X-ray  luminosity and  temperature  was  interpreted by   the
presence  of   cuspy dark  matter halos  suggesting an   early formation  epoch (\citealt{navarro}).  It  was also  suggested   that  fossils are not  overluminous in X-rays but  underluminous in  the
optical, being  comparable to  clusters in   mass and  X-ray luminosity,   but  otherwise  possessing  the  richness and    optical  luminosities of  poor  groups (\citealt{proctor11}).  The   missing
baryons could have been expelled from the system, hidden in the intracluster light or were never present, suggesting that fossils formed in baryon deficient regions in space. Various   studies    also
revealed   that  fossils   exhibit   comparatively high $M/L$ ratios  with  respect to  ordinary  systems (\citealt{vikhlinin,yoshioka,khosroshahi07,proctor11}).

Besides the observational effort to understand the formation and evolution   of fossils, several   simulations have been carried out addressing the same questions from  a theoretical point   of  view.
To  trace    the   evolution   of   present    day   fossils    back   to   their   formation    epoch,  \citet{dariush07}     made   use   of   the    Millennium  and    Millennium   gas  simulations
(\citealt{springel05,hartley08}) revealing that at  any   given redshift, fossils   have  assembled   a higher  fraction   of  their final   halo  mass     compared to  non-fossils with similar  X-ray
luminosity. The latter systems assemble  their  halo mass even at   the present day. These simulations   suggest    that fossils   indeed  formed  early  with most  of their   halo  mass in  place  at
higher  redshifts. Using self  consistent N-body/hydrodynamical    simulations based   on $\Lambda$CDM    cosmology, \citet{donghia05}    found  a   similar result,  a clear correlation between    the
$\Delta    m_{1,2} \ge  2.0$  magnitude   gap   and  the   formation   time  of  groups.  The  earlier  a  galaxy group  is     assembled, the   larger  the magnitude   gap  at $z=0$,  suggesting that
fossils    have  already assembled  half  of  their  final dark  matter    mass  at  $z  \apprge   1$   and    subsequently typically   grow  by     minor mergers  only.    This  early   mass assembly
leaves    sufficient   time for   $L^{*}$    galaxies to  be   first    tidally stripped  and  finally  merge   into  the   central  elliptical by dynamical  friction,  resulting   in the  exceptional
magnitude   gap  at $z=0$  while in   non-fossils the  final dark matter  halo   mass is  assembled quite recently.

The overall motivation driving all  these various observational and theoretical  efforts aim at understanding whether  fossils are a distinct class  of objects resulting from a different formation and
evolution history or being  simply extreme  examples of ordinary groups and clusters. The failure to give  a clear explanation on the true nature of these aggregates is mainly due to the small  number
of objects studied so far, mostly lacking  spectroscopy of group members  for group membership confirmation and a detailed kinematical analysis.

To complement the scarce sample of spectroscopically studied fossils  down to their faint galaxy populations,  the fossil candidate   RX~J1548.9+0851 is studied in detail in this work using  VIMOS MOS
spectroscopy. The system was  selected from the  sample  of \citet{santos} based on its redshift  and the  high  apparent  brightness of  the  central  elliptical. Assuming  an X-ray   temperature  of
$kT =  2$ keV  and  a   metallicity of  $Z=0.4\, Z_{\odot}$, \citet{santos}  estimate an  X-ray luminosity of  $L_{X}=5.09  \times 10^{42}$  ergs      s$^{-1}$  for  the system from   $ROSAT$ count-rates.
At  a redshift of  $z=0.0721$,  the  system   lies    at a     luminosity distance   of    $D_{L}=326$  Mpc and     an angular diameter  distance    of   $D_{A}=283$ Mpc corresponding  to a  spatial
scale of 1.374~kpc~arcsec$^{-1}$.  At this redshift the age of the universe amounts to $\sim   12.5$ Gyr.

The paper is organized as   follows.  In \S 2   observations   and data  analysis techniques   are  described while  \S   3 presents all    obtained results. \S 4  discusses and  compares our findings
with other   fossils   from    the  literature.  Conclusions are given in \S 5.   Magnitudes presented    in    this    work are     SDSS  DR7  model    magnitudes.   Throughout   the    paper,    the
standard  $\Lambda$CDM cosmology   with $\Omega_{M}=0.3$, $\Omega_{\Lambda}=0.7$, and  a  Hubble constant of  $H_{0}=70$ km~s$^{-1}$ is used.

\section{Observations and data reduction}
RX~J1548.9+0851   was observed   with   the  multi-object spectrograph    VIMOS mounted  on the  ESO VLT    UT3  in service mode.    The observations were carried  out in 2009 on February 25  for  the
pre-imaging and on  April 24 and  May 26 for  the associated   multi-object  spectroscopy. VIMOS is   made  up of     four individual  quadrants,  each  with    a   field   of  view  of $\sim7\times8$
arcmin separated  by  gaps   of  2   arcmin. The CCDs  cover  4$\times$2048$\times$4096   pixels with a      pixel  size  of   $15 \times 15  \mu$m  resulting in a  spatial      scale   of       0.205
arcsec pixel$^{-1}$. Considering the  redshift  of the system,  one quadrant comprises roughly  a radius of  300~kpc on the sky. Thus, to cover   essentially the   inner   regions  of  the   group,  a
single pointing      with   the   dominating   elliptical     centered     on VIMOS  quadrant    3    was  scheduled. The pre-imaging  was performed   in the Bessel  $R$  band and consisted   of   one
$200\,$s exposure   carried out   under clear   sky conditions  with a     measured seeing     of   $\sim   0.7$    arcsec  FWHM on    the    images.  For  the   further   photometric analysis,    the
reduced    data  automatically preprocessed     by  the    ESO  reduction     pipeline\footnote{The pipeline      does  bias      subtraction,  overscan     removal,   flat-fielding   and  cosmic  ray
removal.     Images are  photometrically   calibrated   with   photometric  zeropoints,   extinction     coefficients,    airmass,    and  an  estimate of   the sky    background  stored in  the image
header  of  each  individual  quadrant.}    was  considered.   ESO   photometric zeropoints  were  used    to transform    instrumental   magnitudes     to  the   standard   magnitude   system.  Since
zeropoints  are   given   in  ADU~s$^{-1}$, the   reduced  science frames  were  first     divided  by their   exposure    time.  The  resulting magnitudes  were   checked   by    comparing magnitudes
of   some  bright  stars  in   each  quadrant with    SDSS  $r'$    band model     magnitudes.  To   obtain  reliable         radial       velocity      measurements,       the    VIMOS HR        blue
grism          with  a    spectral  resolution     of 0.51$\,$\AA$\,$pixel$^{-1}$    covering   a   wavelength    range   of    4150        to  6200$\,$\AA~was  used.    Spectroscopic    targets  were
selected      based    on    all     SDSS sources   in the      VIMOS    field   brighter  than    $i^{\prime}=20$  mag   that   have    been  photometrically  classified     as    galaxies.   Obvious
stars             falsely classified       as galaxies      as well       as objects     in   the immediate  vicinity        of   the     central  elliptical hardly     seen       on  the    pre-image
have  not       been considered for   the target selection.   To  cover     as    many  faint  galaxies  in    the   inner   region  of  RX~J1548.9+0851   as   possible,   two     slit masks      have
been   prepared  for  quadrant 3.  For brighter objects with  larger spatial extent, slit-lengths     were increased  to    a  more  suitable  value.   The spectroscopic   observations were    carried
out  under clear   sky conditions  and  consisted of 4500   s  exposures for every  slit mask.  The measured  seeing  in the  spectra was  around $\sim  0.8$  arcsec. In   total,  97    slits     were
assigned in  the whole observed  VIMOS      field, out of   which 90  spectra   have been   extracted   by   the  VIMOS    pipeline\footnote{The  method  of  \citet{horne86} is used         to extract
one-dimensional spectra.      The pipeline     applies basic     reduction   steps  (bias   subtraction  and    wavelength  calibration)  and    flux-calibrates  the     raw   spectra.      The   flux
calibration    is     relative  instead     of   absolute,  fluxes being  valid except    for a  constant     factor.    A    median    sky   level    is   estimated for      each   wavelength     and
subtracted    from    the   data.     The  extracted      spectra are    resampled   to        a constant   wavelength  interval     (0.603     \AA~pix$^{-1}$   for       the        HR  blue grism).}.
To    check     for   remaining     sky  lines     in   the pipeline processed     spectra,  each  spectrum was   compared  with    the  associated   sky model.    If a  strong  feature  was   present
at  a  certain   wavelength in  the   model   and   the object spectrum,   it     was deleted from  the object    spectrum manually within  the {\tt   IRAF splot} task.

To   determine   radial   velocities     and      group memberships       of   the      RX~J1548.9+0851 candidate      faint  galaxy    population,   the cross-correlation  technique as  described  by
\citet{tonrydavis}  was applied by    means  of the   {\tt  xcsao} task   in the   {\tt   IRAF  rvsao} package. The analysis yielded redshifts for 84 targets ($\sim 93\%$  of  all extracted  spectra).
Stellar  templates  of   zero   radial  velocity     were taken    from     the  library    of    stellar    spectra  by   \citet{jacoby84}.  Spectral  types similar    to  the  stellar populations of
ellipticals,   i.e.\ G   and    K  giants   have    been  selected     for that  purpose.   To   quantify  the     reliability   of   the   redshift   measurements,   the   confidence  parameter   $R$
provided  by {\tt   xcsao}   has    been   considered.  \citet{kurtzmink98}    have   calibrated   $R$    empirically     and   state that     the   redshift determination is  reliable  above  $R_{\rm
{min}}=3.0$. The   final adopted    redshifts    were   estimated as    the average  from all   utilised stellar  templates while     redshift  errors  were   estimated   from the  associated standard
deviations. Ultimately, redshifts were transformed to  radial velocities by  means of the  relativistic Doppler effect.

\section{The RX J1548.9+0851 system}
\begin{figure}
\centering
\includegraphics[width=\columnwidth]{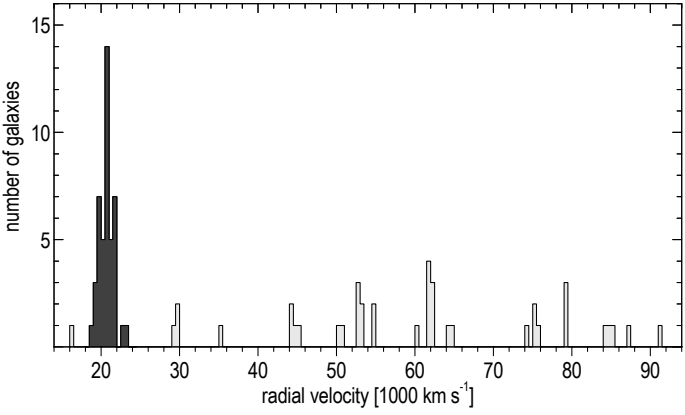}
\caption[Radial  velocity   distribution]{\label{velocityhistogram}Distribution   of    measured   radial    velocities    for   all    galaxies   in the  observed  VIMOS field.    The velocity  range
covers  80000 km  s$^{-1}$  with  a   bin  size of  500   km  s$^{-1}$. Dark  histograms indicate  the RX~J1548.9+0851 system.}
\end{figure}

Fig.~\ref{velocityhistogram}  shows the    distribution  of  radial   velocities   for    all  84 galaxies with    measured redshifts in  the  observed   VIMOS  field.  The  RX  J1548.9+0851 group  is
well-defined in     radial-velocity   space  exhibiting  a    strong  peak  around  $v_{r}\sim  21000\,$km  s$^{-1}$   (dark histograms), confirming   that   it  constitutes   a  real  gravitatioanlly
bound  system.  A    galaxy     with  a  radial   velocity  $v_{i}$    was  considered   a  group     member    if     the      condition        $\left|v_{i}-v_{\rm{sys}}\right|<\Delta    v$        is
fulfilled. According   to \citet{ramella94},  a $\Delta   v$    of  1500   km     s$^{-1}$ is    a  good    match  to   the   scale    of  loose    groups   and  was   used     in   this work.    The
median radial velocity   of  the dark  histograms in   Fig.~\ref{velocityhistogram}  ($v_{\textrm{\scriptsize med}} =  v_{\textrm{\scriptsize sys}} =  20696\,$km  s$^{-1}$)  was   considered as system
velocity. The thus defined  sample of  group members holds until  a   $\Delta v$ of 1250   km   s$^{  -1}$ is applied\footnote{Accounting    also   for radial  velocity error   bars.}.   

\begin{figure}
\centering
\includegraphics[width=\columnwidth]{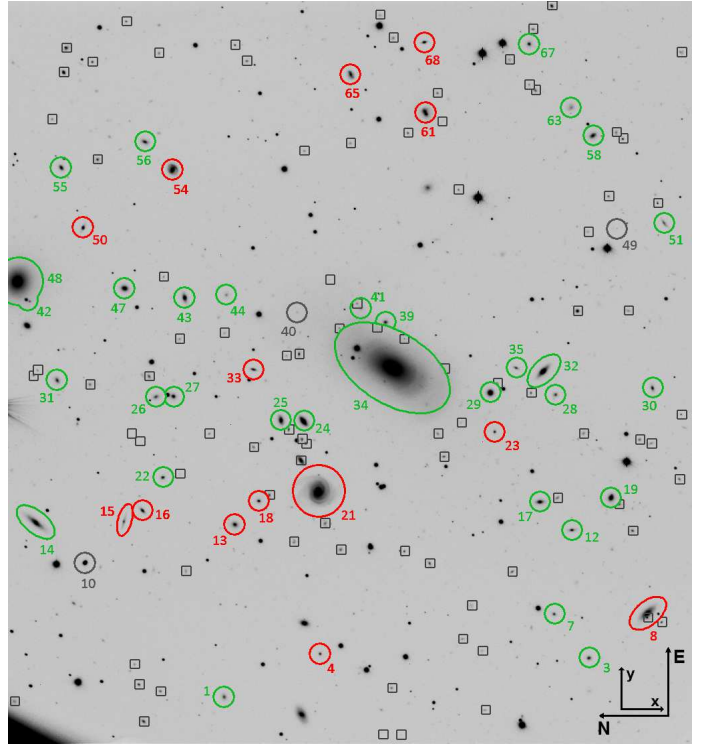}
\caption{\label{q3}Spectroscopic targets  in the  central region  of the  RX~J1548.9+0851 system.  Green circles  indicate group  members while  red ones   are background   galaxies. Grey  circles are
spectroscopic targets without  a reliable redshift.  Squares show photometrically  classified galaxies in  the  SDSS down  to $i'=21$ mag.  The field-of-view is  $\sim7\times 8$ arcmin (VIMOS quadrant
3 -- orientation following  the   CCD mask  coordinates ($x,y$).  The alignment of  equatorial    coordinates ($N,E$) is also  shown). Numbers refer to the ID of Table \ref{members}.}
\end{figure}

\begin{figure}
\centering
\includegraphics[width=\columnwidth]{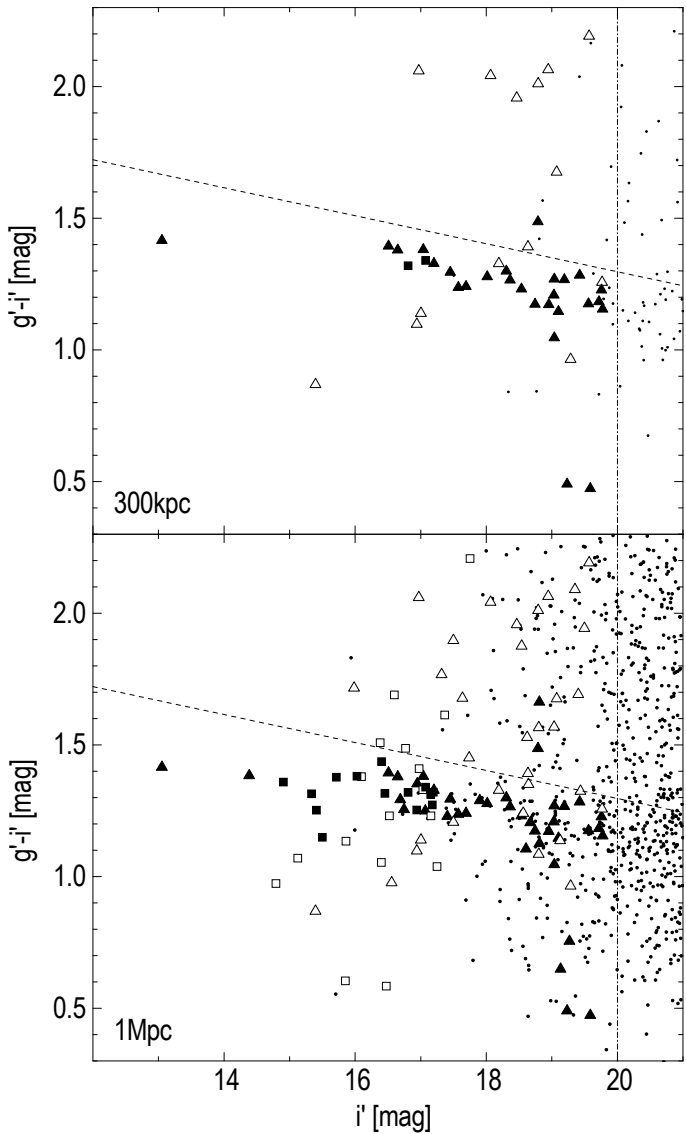}
\caption{\label{cmds}   Colour-magnitude diagrams for   all galaxies within 300$\,$kpc   and 1$\,$Mpc. Galaxies with   spectra are shown as  triangles (this work) and squares (literature). Small  dots
are photometrically  classified SDSS galaxies with $i'  <21$. Filled symbols show   spectroscopically confirmed group members  while open ones are  background galaxies. Group members  form a tight red
sequence of $\sim 0.4$~mag thickness . The dashed  lines show the   adopted  upper limit  of this sequence  $g'-i' =  -0.05317\cdot  i'+2.36$. Vertical lines  show the magnitude  limit ($i'=20$ mag)  for the spectroscopic  target
selection.}
\end{figure}

Out of  the   84  measured redshifts,   41 ($\sim49\%$)   have   been   identified as   group  members  in  the  whole VIMOS     field, amounting  to 40 identified members   within 1     Mpc  of   the
central     elliptical. Complementing  the present measurements,  31 additional redshifts   have  been found   in the literature, resulting in a total number of 54   spectroscopically confirmed  group
members within 1 Mpc, amounting  to $\sim51\%$ of   all studied  redshifts within this  radius (see  also Fig.\ \ref{cmds}).  Figure~\ref{q3}  shows  the    VIMOS $R$   band pre-image     of  the  inner region
of the RX J1548.9+0851 system.   Green  circles indicate  spectroscopically  confirmed   group   members   while    red  ones   are   background   galaxies. Squares show   photometrically classified    galaxies in   the    SDSS   down  to   $i'=21$
mag.   The inner  region of the system is dominated by  two bright ellipticals instead  of  only one as expected for a fossil group. The second brightest  elliptical is located at a   projected distance
of 326  kpc from the central galaxy.  With  a magnitude difference of $\Delta  m_{1,2}=1.34$ in  the $r'$   band,  it clearly violates the 2   mag  criterion from \citet{jones03} leaving the question of
RX J1548.9+0851 being a fossil  or not to the definition  of  the virial radius.  

Table     \ref{members} lists,  amongst others,     the  radial     velocities    and   associated   confidence  parameters    $R$    of   the   studied  group     population  within    1 Mpc   of the
central elliptical.    Galaxies are   listed    with  increasing  right    ascension,   i.e.\ from bottom  to   top in Fig.~\ref{q3}. 

\subsection{The colour-magnitude diagram}
Figure \ref{cmds}   shows    colour-magnitude  diagrams of    all SDSS  galaxies   down  to  $i'    = 21$ mag   within  300 kpc   and  1 Mpc   of  the central elliptical.  The  magnitude limit of  the
spectroscopic    target    selection    is  indicated   as   vertical  line. At  a  luminosity  distance of   $D_{L}=326$~Mpc    based on  the median  redshift  of  the system,  this magnitude   limit
corresponds to an   absolute   magnitude  of $M_{i'}\sim -16.9$  mag. Objects     with     untypical  magnitudes   and    colours    lying     outside   the    main    point  distribution   have  been
checked  in   SDSS.   Most of  these   objects   were identified  as  stars  misclassified as   galaxies   and  were removed from    the  plot. Small dots are all SDSS  sources  in  the field brighter
than $i=21$ mag, photometrically  classified as galaxies.  Triangles show the  measured redshifts  from   this work while squares  indicate the   additional redshifts found  in  SDSS  and NED.  Filled
symbols  show  confirmed  group  members   while  open ones  are   background  galaxies. The  diagrams   clearly  reveal the   increasing incompleteness   of  redshift  measurements towards    fainter
magnitudes.  The faintest    galaxy  with  a  reliable    redshift  in  the investigated sample    has a  magnitude of    $i' = 19.78$~mag.   Also the spatial incompleteness   towards larger radii is
evident. While   the spectroscopically studied sample  within 300 kpc  is well represented  above  $i'  =  20$ mag, having  68\% measured redshifts  compared to  all  SDSS galaxies in the field,  this
fraction drops to  24\% within 1 Mpc.   Nearly all redshifts  from  the literature are found  in the outer regions   of the system,   where no VIMOS measurements were taken. Group members    form   an
unambiguous  red    sequence   from   the    brightest   galaxy   into    the  dwarf  regime     while  non-members    are   only  diffusely  distributed   in   the  diagrams. The  dashed  lines    in
Fig.$\,$\ref{cmds}   indicate   the adopted  red   sequence  upper  limit.  Only    2 members    are  found   above   this    upper limit. Interestingly, these  2 galaxies   (39 and   42)  are located
in   the immediate vicinity  of the two brightest ellipticals   in the      group (see   Fig.$\,$\ref{q3}).  The identified red sequence of spectroscopically confirmed group members is well-defined  in
colour and shows a  lack of galaxies that are  bluer than the adopted red sequence  upper  limit by $\ge$  0.4 mag. Galaxies  bluer than this  limit are at  least $\sim$6 magnitudes  fainter  than the
central elliptical and comprise only 4 galaxies in the whole spectroscopic sample, confirming that the RX~J1548.9+0851 system is dominated by old and passively evolving galaxies.

\begin{figure}
\centering
\includegraphics[width=\columnwidth]{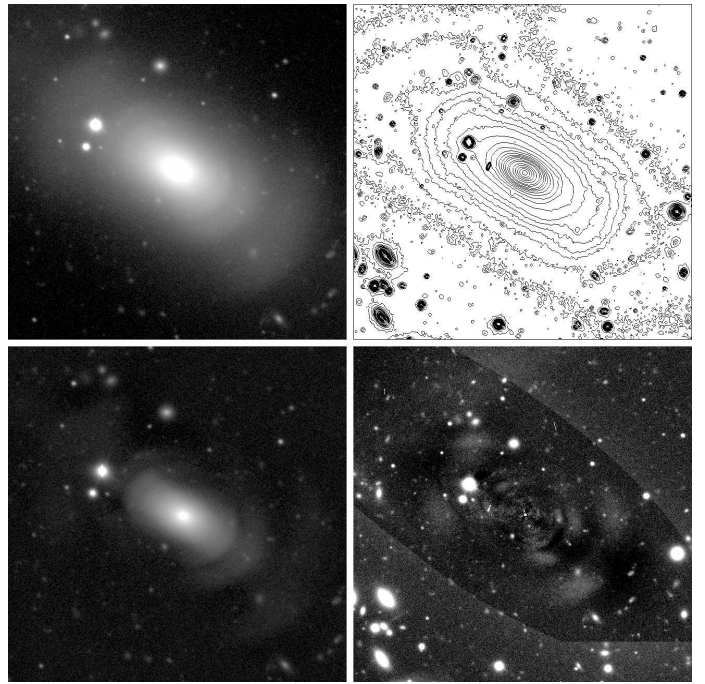}
\caption{\label{centralelliptical}Morphology of  the  central  elliptical.   Upper  left: $R$   band  image,  upper right:   isophote map,  lower   left: $R$  band  image  after  subtraction   of  the
median box  filter,  lower right:  residuals  to  the  {\tt IRAF ellipse}  model. Shells are   clearly visible in  the  lower  panels  and show strong   symmetry along the   major axis. The   isophote
map reveals boxy isophotes  in the  outer regions of the galaxy, resulting from the observed shell structure.  The residuals  illustrate the  large contamination of foreground  sources which have been
masked within the {\tt   ellipse} task. The field of view is $\sim 1.5$ arcmin on a side in the left panels and $\sim 2.2$ arcmin in the right ones.}
\end{figure}

\subsection{The central elliptical}
\begin{figure}
\centering
\includegraphics[width=\columnwidth]{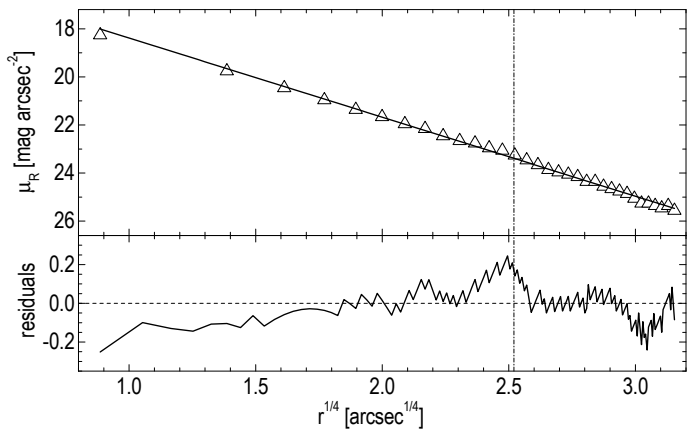}
\caption{\label{devaucouleurs}   Surface  brightness  profile  of  the  central   elliptical plotted  against $r^{1/4}$.  The straight  line  is  the deVaucouleurs  fit to the data.  The vertical line
indicates the effective  radius. The lower  panel shows the residuals to the fit. No  light excess characteristic for a cD galaxy is seen.}
\end{figure}

Since the   central   elliptical  in  fossil    groups  is  supposed    to   contain  the    merger  history  of    the progenitor group  population,  its  morphological  appearance   is of    special
interest.    Figure \ref{centralelliptical}~a shows   the central    elliptical in  the VIMOS   $R$   band pre-image.  Symmetric shells  are  unambiguosly identified  along  the  galaxy  major   axis.
To  enhance   the appearance of   these features,      the image   was smoothed  in  {\tt IRAF}    by   means   of  a    median     box   filter   and subtracted by  this unsharp mask. The   residuals
reveal  the observed   shell structure  in much more   detail (Fig.\ \ref{centralelliptical}    c) and confirm  the strong  symmetry  as  already indicated   by the unaltered frame.  In addition,  the
surface-brightness profile  of  the  central elliptical  has  been  studied with {\tt   IRAF   ellipse}.  Due     to the  galaxy's     large   spatial extent,  its   diffuse   light  distribution   is
strongly contaminated  by  numerous   smaller   objects.  Therefore care    was  taken    in  masking out   all  these   sources    properly   in  the isophote fitting procedure  to avoid   systematic
deviations  in    the  resulting surface  brightness profile.   Ellipticity, position  angle and    central  coordinates  of the   successive ellipses  were allowed   to change   with  radius.  Figure
\ref{centralelliptical}~d shows the   residuals between the   $R$  band image  and the obtained   isophote model, again    revealing the pronounced shell  structure. To  quantify the    shape of   the
surface-brightness  profile, a   deVaucouleurs   $r^{1/4}$  law was  fit   to  the data from   outside the  seeing dominated    center (0.7  arcsec) down    to $\mu_{R}\sim  25$~mag~arcsec$^{-2}$. The
effective radius was estimated to  $r_{e}= 40$~arcsec corresponding to $\sim55$~kpc.   Figure \ref{devaucouleurs} shows  the  profile   in $\mu_{R}$  mag arcsec$^{-2}$ against   $r^{1/4}$.  The  solid
line  is  the best-fit deVaucouleurs  law.  Residuals indicate only  minor deviations from the model in  the order of 0.1~mag.  No light excess above the deVaucouleurs law is seen that would  classify
the elliptical as a cD galaxy.

\subsection{SSP ages and metallicities of group members}
\begin{figure}    
\centering   
\includegraphics[width=\columnwidth]{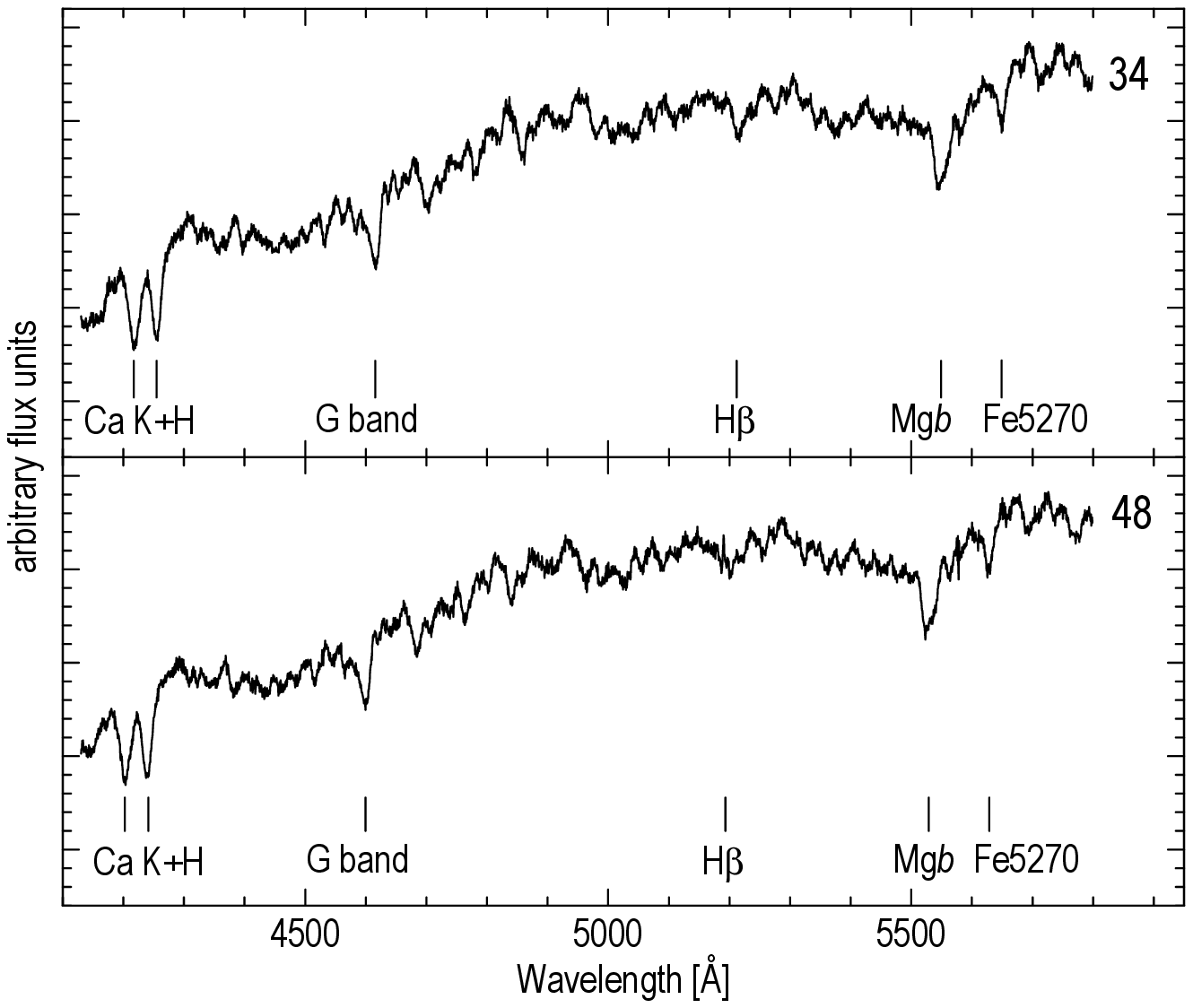}\\
\vspace{0.1cm}
\includegraphics[width=\columnwidth]{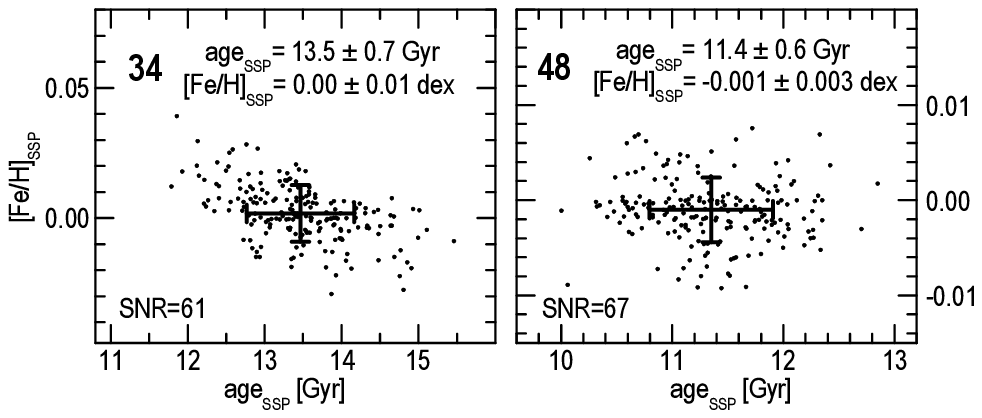}
\caption[Radial velocity  distribution]{\label{ellspec}  Upper  panels: Spectra   of  the  two  brightest  ellipticals in    the RX~J1548.9+0851  system. The  most prominent  absorption  features  are
identified.  As  expected,  the   galaxies  are  dominated  by   an  old  stellar  population.   Lower panels: Monte-Carlo simulations  of the  corresponding SSP  fits.  Crosses indicate averages  and
standard deviations of each point distribution, presenting  the final adoped fitting results.}
\end{figure}

To        determine        ages          and         metallicities          of           the        RX~J1548.9+0851         group           population,           the           open-source      package
ULySS\footnote{\url{http://ulyss.univ-lyon1.fr/}}   ({\bf    U}niversity      of       {\bf  Ly}on       {\bf  S}pectroscopic   Analysis       {\bf  S}oftware; \citealt{koleva09})   was used  to   fit
synthetic  single     stellar  population (SSP)   models based    on  a    library      of stars with    varying   atmospheric  parameters   directly  to     the observed    galaxy  spectra   yielding
SSP  equivalent  ages  and  metallicities  of the stellar  populations  of the investigated  galaxy  sample.     ULySS  makes    use of  the       Pegase$\,$HR     models   from     \citet{leborgne04}
based      on      the Elodie$\,$3.1  library  of stellar         spectra (\citealt{elodie3,elodie31}). Since  the   synthesis   of a    stellar population   requires   a stellar  spectrum   at    any
point   in      the  parameter    space  ($T_{\rm eff}$,   log~$g$    and     [Fe/H]),      the  library    involves    an  interpolator  for       that purpose.    All  Pegase$\,$HR   models   are
computed   assuming  a   Salpeter     IMF     and  Padova  1994   evolutionary     tracks  providing   synthetic  SSPs     with   ages  between  1$-$20000$\,$Myr    and  metallicities  between  $-2.3
-0.69\,$dex.   A multiplicative  polynomial is used  to adjust the overall spectral shape to   the SSP   model.  The order of the  polynomial   can  be freely   chosen by the  user but  shouldn't   be
taken   too  low  resulting   mostly in  a   severe   mismatch of  the investigated  spectrum  and the  SSP   model.  Balancing between a   minimum  order  to achieve  reasonable, stable  values and a
maximum  order to avoid long,  unnecessary computing time, an   order of 40 was chosen for all galaxies. To evaluate the reliability of the  performed SSP fits,  $\chi^{2}$ and convergence maps of the
investigated  age-metallicity  parameter space  have been  considered.  200  Monte-Carlo  simulations  have been  computed for  every galaxy,  repeating a    fit  successively   with   random Gaussian
noise\footnote{The  dimension  of  the  added  noise was based   on  a  user-defined signal-to-noise  ratio provided   for each  galaxy.  The S/N ratios  were measured  with  {\tt    IRAF  splot} at a
rest-frame wavelength     of    $\sim5140\,$\AA.}   applied  to   the  spectrum  in    each  step. The  resulting   point    distributions  were     then used    to  calculate   average  SSP  ages and
metallicities  and  the associated standard deviations (excluding  obvious outliers).   Figure \ref{ellspec}  shows  the spectra  of  the   two brightest  ellipticals in  the RX~J1548.9+0851    system
and  the resulting Monte-Carlo simulations  of the corresponding  SSP fits. Table   \ref{members} lists the  measured SSP  equivalent ages and  metallicities of all RX~J1548.9+0851      group  members
within 1 Mpc.

\begin{figure}
\centering
\includegraphics[width=\columnwidth]{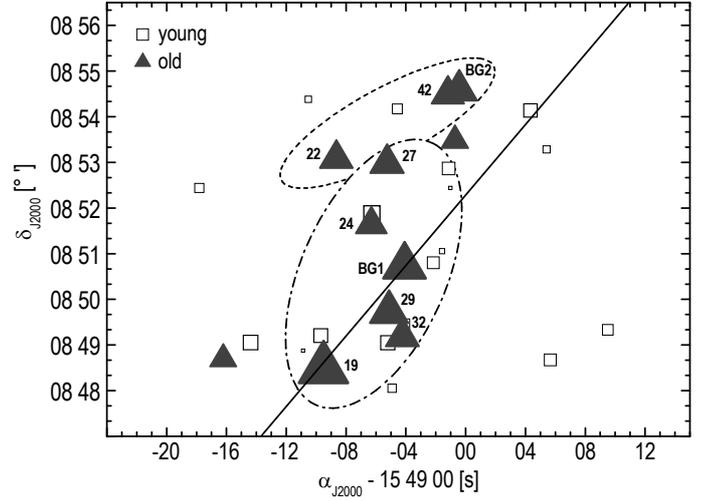}
\caption{\label{agemetallicitymaps}Age  map   of the  inner  $\sim\,$300$\,$kpc  region  of  RX~J1548.9+0851 (see  Fig. \ref{q3}).   There is  a  clear   segregation in age  for   the
investigated  faint  galaxy  population.  While  younger objects  are  diffusely  distributed  and  found mostly  in the outskirts  of the system, the oldest objects ($>$ 8 Gyr) form  a clear  elongated
structure aligned   with the orientation of the major axis of  the  central elliptical, indicated as dashed line. Ellipses contain the same objects as those in Fig.\ \ref{agevelocity}. See text  for
details.}
\end{figure}

To   investigate   the   correlation of   the   measured   SSP  ages      with  the   spatial   distribution   of  member galaxies, Fig. \ref{agemetallicitymaps} shows  ages of galaxies in  the  inner
$\sim\,$300$\,$kpc  region  of the  RX~J1548.9+0851  system. In  Fig.\  \ref{agemetallicitymaps} a, galaxies  have been separated   into old  and  young objects at  an arbitrary age  threshold  of   8
Gyr. There is a  clear segregation in  age for   the investigated   galaxy population. While  younger objects are  diffusely  distributed and found  mostly in  the outskirts of the system,  the oldest
objects form   a clear elongated structure  aligned  with the  orientation  of  the major  axis of the    central  elliptical, indicated  as  dashed  line.  

\subsection{Galaxy velocity distribution}
\begin{figure}
\centering
\includegraphics[width=\columnwidth]{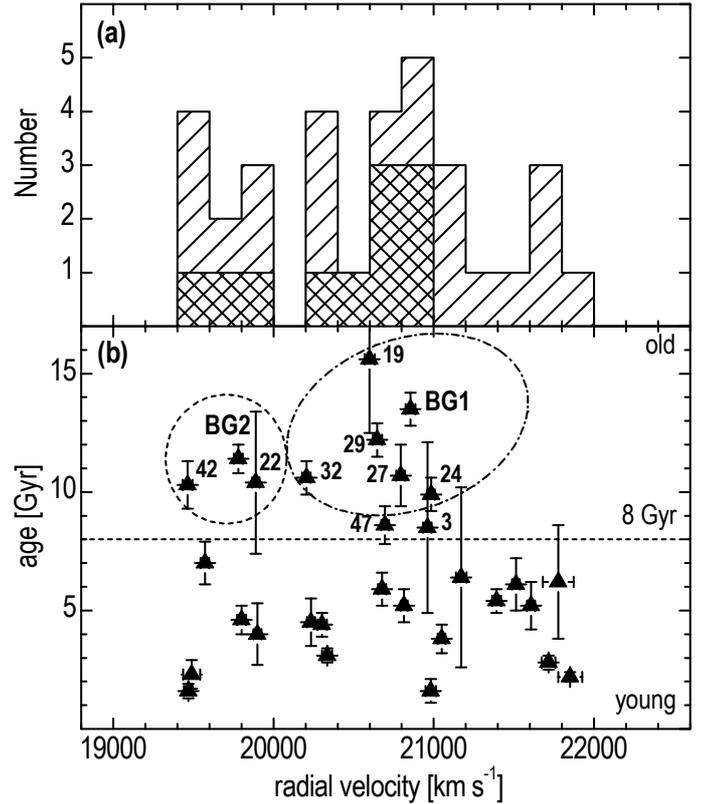} \caption{\label{agevelocity}   Galaxy  velocity   distribution  of  the inner   $\sim\,$300$\,$kpc region  of the RX~J1548.9+0851  system ($N=32$)
correlated with  the measured   SSP   ages (a):   Galaxy  velocity     distribution. Old  galaxies ($>$   8 Gyr) are   indicated   as cross-hatched  histograms. (b): Age$-$velocity plane.  Galaxies  are
separated into   old  and  young objects.  Ellipses contain the same objects  as those in  Fig.\ \ref{agemetallicitymaps} and  separate  old galaxies   into subsamples around  BG1  and BG2.        See
text for details.}
\end{figure}

Figure   \ref{agevelocity}$\,$a  shows   the   galaxy   velocity   distribution     of  spectroscopically   confirmed  group  members    in  the     inner  $\sim\,$300$\,$kpc  of the   system ($N=32$)
while   Fig.   \ref{agevelocity}$\,$b    relates     the    measured   radial   velocities   with   the    corresponding  SSP     ages.  Based   on   the   spatial  segregation   in  age   found    in
Fig.$\,$\ref{agemetallicitymaps}, galaxies  are again separated into \emph{old}   and \emph{young} objects at   the arbitrary  age  of   8 Gyr.    The two brightest  ellipticals,   hereafter  BG1  and
BG2,  are  the oldest objects  in the sample  (with  the exception  of   one  faint    galaxy,   exhibiting    a   huge    error      bar  in     age,  however)     and  build      up  the centers  of
two subsamples  in the  velocity  distribution of  old galaxies  (cross-hatched  histograms).  While the   oldest galaxies  around  BG1 form the   core   of the overall    velocity distribution,   old
galaxies   around BG2   are  found at   systematically  lower velocities. Hence, the  luminosity-weighted mean  radial velocity  $v_{r}=20633$ km  s$^{-1}$ is  significantly   lower  than  the  radial
velocity  of the  central elliptical $v_{r}=20854$ km s$^{-1}$.  In  contrast to  the old  galaxy  population, young objects       show a smooth   distribution  over   all   velocities and exhibit   a
much larger    dispersion. Dashed   ellipses   in  Fig.$\,$\ref{agemetallicitymaps}   and  \ref{agevelocity}$\,$b  contain  the   same objects,  indicating that the subpopulations    around  BG1  and
BG2 in  radial  velocity  space  are  also more or  less spatially correlated.

\subsection{Velocity dispersion}
To evaluate the dynamical properties  of the RX~J1548.9+0851   system,  the  luminosity weighted dynamical formulae as  presented  in  \citet{firth06} have   been   applied.  Table    \ref{dispersion}
lists     the    luminosity-weighted and non-weighted velocity   dispersions for  all spectroscopically  confirmed group members  within 1  Mpc and for    the      sample  of the inner   region  shown
in Fig.$\,$\ref{agemetallicitymaps}  and   \ref{agevelocity}.    Dispersions have   also been    calculated excluding   young galaxies  and   BG2.    Focusing  on all   spectroscopically
confirmed members within 1    Mpc  leads to   the   largest  velocity   dispersion  (568   km    s$^{-1}$)   while  restricting  to  the   inner   $\sim\,$300$\,$kpc    region  leads to    a  slightly
smaller value (496  km    s$^{-1}$).  The old population  within $\sim\,$300$\,$kpc   yields a  value of (440   km    s$^{-1}$) . This     restriction     does  not   lower  the
measured dispersion as  severely  as the    exclusion of BG2  which    dominates  the  luminosity  of    all other   faint and    young   galaxies   by  far.  Omitting   BG2  leads to   much   smaller
values in   the  central region for both    the whole     (321  km      s$^{-1}$)     and     old      (266   km    s$^{-1}$)        galaxy    populations. Non-weighted dispersions show  significantly
higher values  because  of  the large   spread in  velocity  of  young, faint   galaxies.  The presented    results    are      in    agreement  with the   study   of  \citet{mendesdeoliveira06}   who
found that emission-line   galaxies     primarily  populate      the  outskirts     of   the   galaxy   velocity     distribution of   the   fossil  cluster RX~J1552.2+2013     suggesting that  young,
star-forming galaxies have recently been accreted by the  group and are not yet virialised.

\subsection{Dynamical mass}
\begin{figure}
\centering
\includegraphics[width=\columnwidth]{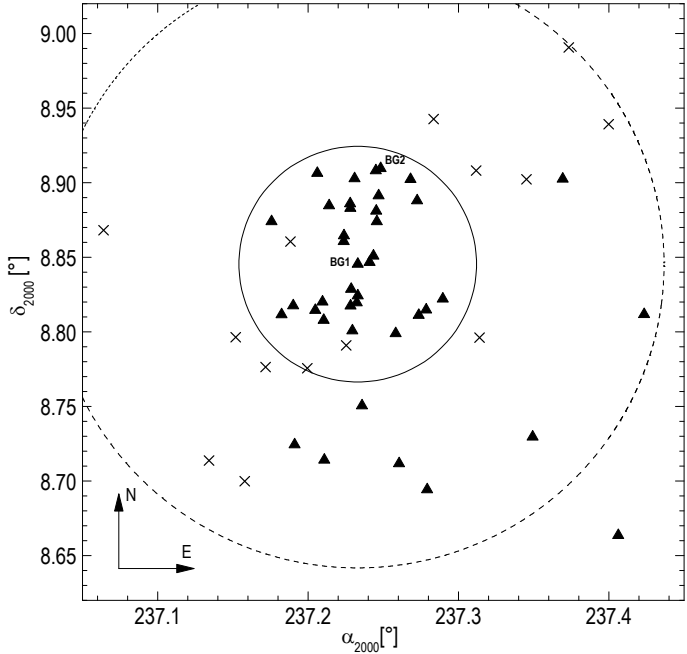}
\caption{\label{group}Spatial distribution of   spectroscopically confirmed group members  from this work (triangles)  and the literature (crosses).  The inner circle shows  the luminosity-weighted
harmonic mean radius centred on the dominating elliptical. The dashed circle indicates a radius of 1 Mpc. The two brightes ellipticals are marked as BG1 and BG2.}
\end{figure}

Table \ref{lumdynamics} summarizes the luminosity-weighted dynamical  properties  of RX~J1548.9+0851 derived from  all spectroscopically  confirmed group members within 1  Mpc. Weightings  $w_{i}$ are
based on the  apparent SDSS $r^{\prime}$ band  magnitudes of group member galaxies,  converted to relative luminosities. The dynamical mass  of   RX~J1548.9+0851 was  determined using   two  different
mass estimators,   i.e.\  virial and  projected   mass. For  that  purpose,  the harmonic  mean  and   virial  radii have  been  calculated. The computation   of the harmonic mean radius  $R_{H}$ (see
Fig.\ \ref{group}) involves the determination of  the distances  $\left|{r_{i}-r_{j}}\right|$ of  all  $N(N-1)/2=1431$ galaxy pairs within the  sample.  For the  projected mass,  the   projected radial
distances  $r_{\bot  i}$ from  the central  elliptical  have  been applied.   Characteristic  system  crossing times   were estimated by    dividing  the average  projected distance of  group  members
from  the central elliptical by  the average speed of group  members relative to the  group  centre of mass,  considered as the central  elliptical.  Finally, the mean    mass from both estimates   is
given with  the  corresponding  standard deviation   as error  estimate. As  already expected  by the  relatively large  velocity dispersions  as shown  in  Table  \ref{dispersion} the  system shows a
comparatively large  dynamical mass,  typical for   a cluster  rather   than  a   group,     in agreement   with   previous work   (\citealt{mendesdeoliveira06,mendesdeoliveira09,cypriano,proctor11}).
The crossing    times below   1 Gyr  for the  inner $\sim\,$300$\,$kpc of RX~J1548.9+0851      suggest that members had enough  time to virialize within the group potential.

\begin{table}
\begin{minipage}[t]{\columnwidth}
\caption{\label{dispersion}Velocity dispersion measurements.}
\centering
\renewcommand{\footnoterule}{}
\begin            {tabular*}{\columnwidth}{@{\extracolsep{\fill}}p{2cm}p{0.7cm}p{0.7cm}p{0.7cm}p{0.7cm}p{0.7cm}}
\hline
                                                                                                                                                                                    &   \multicolumn{1}{c}{(1)}      &      \multicolumn{1}{c}{(2)}     &     \multicolumn{1}{c}{(3)}      &       \multicolumn{1}{c}{(4)}    &  \multicolumn{1}{c}{(5)}    \\
\hline                                                                                                                                                                                                                                                                                                                                                        
           \multicolumn{1}{c}{$N$}                                                                                                                                                  &   \multicolumn{1}{c}{54}       &      \multicolumn{1}{c}{32}      &     \multicolumn{1}{c}{31}       &     \multicolumn{1}{c}{11}       &  \multicolumn{1}{c}{10}     \\
  \multicolumn{1}{c}{$\sigma              =  {\left[ {\frac{1}{N}\sum\nolimits_i {\Delta {v_i}^2} } \right]^{\frac{1}{2}}} $ }                                                      &   \multicolumn{1}{c}{693}      &      \multicolumn{1}{c}{720}     &     \multicolumn{1}{c}{716}      &     \multicolumn{1}{c}{499}      &  \multicolumn{1}{c}{475}    \\
  \multicolumn{1}{c}{$\sigma_{\rm{lum}}  = {\left[ {\frac{{\sum\nolimits_i {{w_i}{{\left( {\Delta {v_i}} \right)}^2}} }}{{\sum\nolimits_i {{w_i}} }}} \right]^{\frac{1}{2}}}$ }     &   \multicolumn{1}{c}{568}      &      \multicolumn{1}{c}{496}     &     \multicolumn{1}{c}{321}      &     \multicolumn{1}{c}{440}      &  \multicolumn{1}{c}{266}    \\
\hline            
\end{tabular*}     
\begin{footnotesize}
\begin{flushleft}
{\bf  Notes:}  All values in km s$^{-1}$.  \\
$(1)$:  Calculations for all spectroscopically confirmed members  within 1 Mpc including redshifts from literature ($N=54$).                        \\
$(2)$:  Calculations for all members within $\simeq300$kpc -- sample as shown in Figs.\ \ref{agemetallicitymaps} and \ref{agevelocity} ($N=32$).     \\
$(3)$:  Same as (2) excluding BG2 ($N=31$).                                                                                                          \\
$(4)$:  Restricting to old ($>8$Gyr) galaxies ($N=11$).                                                   \\
$(5)$:  Same as (4) excluding BG2 ($N=10$).
\end{flushleft}
\end{footnotesize}
\end{minipage}
\end              {table}

\subsection{The galaxy luminosity function}
\begin{figure*}
\centering
\includegraphics[width=500pt]{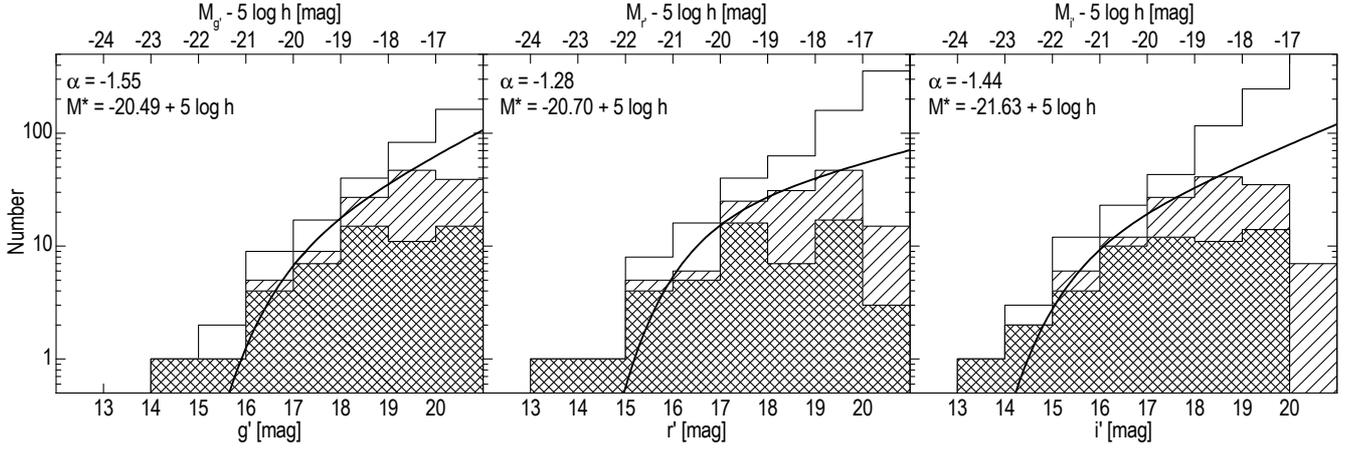}
\caption{\label{schechter}The   galaxy  luminosity     function   of the    RX~J1548.9+0851   system  ind  the  SDSS $g^{\prime}$,     $r^{\prime}$, and $i^{\prime}$  bands. Open  histograms  indicate
all    SDSS galaxies  within    1  Mpc  while  hatched  ones show  photometric members  (see  text) and cross-hatched  ones the spectroscopically confirmed group sample. Solid lines  are the  best-fit
Schechter functions. The central elliptical  has been excluded in the fits. The Schechter fit parameters $\alpha$ and $M^{*}$ are shown for each passband.}
\end{figure*}

Following  previous   work  on fossils,  the  shape  of  the  galaxy  luminosity  function of  RX~J1548.9+0851  has  been studied     in
the  SDSS $g^{\prime}$,   $r^{\prime}$,  and      $i^{\prime}$  bands.    To correct  for   incompleteness in  the  sample of   spectroscopically determined   group members,    photometric    redshift
models  have  been adopted   from the    SDSS\footnote{Three    photometric redshift    models are   stored  in    the   SDSS.   To evaluate     the accuracy   of   these   models, the  derived  VIMOS
redshifts have  been compared with   the corresponding    photometric  redshift  data  from  the SDSS.   Based on   this  comparison,  redshift thresholds for group membership have  been   defined  in
all three  models. A galaxy was  considered  a  photometric  group  member  only   if it  was   considered  as  such   in all  three models.}   for all  galaxies with   no  observed spectrum.    These
additional photometric redshifts  confirmed the     clear segregation   of    galaxies   into     members   and      non-members at    the     red   sequence  upper  limit.  Figure     \ref{schechter}
shows    the    galaxy luminosity      distribution  of RX~J1548.9+0851  member galaxies   in    the   SDSS   $g^{\prime}$, $r^{\prime}$,     and     $i^{\prime}$  bands within 1  Mpc of  the  central
elliptical.  Open   histograms show  all  SDSS  galaxies    while  cross-hatched  ones    refer     to    spectroscopically   confirmed   members.    Dashed     histograms    indicate  photometrically
determined group   members  and evidently dominate  the    studied  group  population    at fainter    magnitudes.  Solid    lines   show   the corresponding    best-fit Schechter    functions.    The
central     elliptical  was     excluded   in      this procedure.     In  all     three    passbands,      the derived  luminosity  functions    exhibit   a  steep faint    end  ($\alpha_{g'}=-1.55$,
$\alpha_{r'}=-1.28$,  $\alpha_{i'}=-1.44$)   resulting in   an     average slope  of   $\alpha=-1.4\pm0.1$.  The   overall   shape of the   luminosity   function doesn't  reveal any peculiarities.  No
pronounced dip is visible at  $M_{i^{\prime}}=    -18+5 \log h$ as has been  observed  in the fossil RX J1552.2+2013 (\citealt{mendesdeoliveira06}).

\begin{table}
\begin{minipage}[t]{\columnwidth}
\caption{\label{lumdynamics}Luminosity weighted dynamical properties of the RX~J1548.9+0851 system.}
\centering
\renewcommand{\footnoterule}{}
\begin            {tabular*}{\columnwidth}{@{\extracolsep{\fill}}p{2.6cm}p{3.5cm}p{1.5cm}}

\hline
                \multicolumn{1}{c}{(1)}            &                   \multicolumn{1}{c}{(2)}                                                                                                                                                                                                        &  \multicolumn{1}{c}{(3)}        \\ 
\hline                                                                                                                                                                                                                                                                                                                                     
\\                                                                                                                                                                                                                                                                                                                                         
\rm {Mean~radial~velocity}                         &    $\overline{v}=\frac{{\sum\nolimits_i {{w_i}{v_i}} }}{{\sum\nolimits_i {{w_i}} }}                                                                                                                                                          $   &  \multicolumn{1}{c}{20731 km s$^{-1}$}      \\ 
\rm {Velocity~dispersion}                          &    $\sigma_{\rm{lum}}     ={\left[ {\frac{{\sum\nolimits_i {{w_i}{{\left( {\Delta {v_i}} \right)}^2}} }}{{\sum\nolimits_i {{w_i}} }}} \right]^{\frac{1}{2}}}                                                                                 $   &  \multicolumn{1}{c}{568 km s$^{-1}$}      \\ 
\rm {Harmonic~radius}                              &    $R_{\rm{H}} ={\left[ {\frac{{\sum\nolimits_i {\sum\nolimits_{j < i} {\left( {{w_i}{w_j}} \right)/\left( {{r_i} - {r_j}} \right)} } }}{{\sum\nolimits_i {\sum\nolimits_{j < i} {\left( {{w_i}{w_j}} \right)} } }}} \right]^{-1}}           $   &  \multicolumn{1}{c}{388 kpc }      \\ 
\\                                                                                                                                                                                                                                                                                                                                         
\rm {Virial~radius}                                &    $r_{\rm{V}} =\frac{{\pi {R_{\rm{H}}}}}{2}                                                                                                                                                                                                 $   &  \multicolumn{1}{c}{609 kpc  }      \\ 
\\                                                                                                                                                                                                                                                                                                                                         
\rm {Crossing~time}                                &    $t_{\rm{c}} =\frac{{\left\langle r \right\rangle }}{{\left\langle v \right\rangle }}                                                                                                                                                      $   &  \multicolumn{1}{c}{0.68 Gyr   }      \\ 
\\                                                                                                                                                                                                                                                                                                                                         
\rm {Virial~mass}                                  &    $M_{\rm{V}} =\frac{3}{G}{\sigma ^2}{r_{\rm{V}}}                                                                                                                                                                                           $   &  \multicolumn{1}{c}{$1.37 \times 10^{14} \rm{M}_{\odot}$     }      \\ 
\\                                                                                                                                                                                                                                                                                                                                         
\rm {Projected~mass}                               &    $M_{\rm{p}} =\frac{{32}}{{\pi G}}\frac{{\sum\nolimits_i {{w_i}{{\left( {\Delta {v_i}} \right)}^2}{r_{ \bot i}}} }}{{\sum\nolimits_i {{w_i}} }}                                                                                            $   &  \multicolumn{1}{c}{$3.70 \times 10^{14} \rm{M}_{\odot}$     }      \\ 
\\                                                                                                                                                                                                                                                                                                                                         
\rm {Mean~mass}                              &    $\nicefrac{1}{2}\left(M_{\rm{V}}+M_{\rm{p}}\right)$                                                                                                                                                                                           &  \multicolumn{1}{c}{$2.5 \pm 0.8 \times 10^{14} \rm{M}_{\odot}$      }      \\ 
\hline            
\end{tabular*}     
\begin{footnotesize}
\begin{flushleft}
{\bf  Notes:}  Velocities are given in km s$^{-1}$, radii in kpc, masses in $10^{14}$M$_{\odot}$, and the crossing time in Gyr. \\
$(1)$:  Description.                                                                                                            \\
$(2)$:  Dynamical formulae.                                                                                                     \\
$(3)$:  Calculations for all spectroscopically confirmed members  within 1 Mpc including redshifts from literature ($N=54$).    \\
\end{flushleft}                                                   
\end{footnotesize}
\end{minipage}
\end              {table}                                                                                                       

\begin{table}
\begin{minipage}[t]{\columnwidth}
\caption{\label{masstolight}Group luminosities, luminosity function and mass-to-light ratios.}
\centering
\renewcommand{\footnoterule}{}
\begin            {tabular*}{\columnwidth}{@{\extracolsep{\fill}}p{0.5cm}p{2cm}p{0.7cm}p{0.7cm}p{0.7cm}}
\hline
                        &                                                                                                                 &   \multicolumn{1}{c}{$g^{\prime}$}   &     \multicolumn{1}{c}{$r^{\prime}$}     &     \multicolumn{1}{c}{$i^{\prime}$} \\
\hline                                                                                                                                                                                                                                                               
\multicolumn{1}{c}{(1)} & \multicolumn{1}{c}{${L_{{\rm{tot}}}} = \sum\nolimits_i {{L_i}}$}                                                &   \multicolumn{1}{c}{5.22}           &   \multicolumn{1}{c}{6.35}               &     \multicolumn{1}{c}{7.62}         \\
                                                                                                                                                                                                                                                                   \\
\multicolumn{1}{c}{(2)} & \multicolumn{1}{c}{${L_{{\rm{tot}}}} = \int_0^\infty  {L\Phi (L)dL}$ }                                          &   \multicolumn{1}{c}{6.25}           &   \multicolumn{1}{c}{5.90}               &     \multicolumn{1}{c}{8.41}         \\
                                                                                                                                                                                                                                                                   \\
\multicolumn{1}{c}{(3)} & \multicolumn{1}{c}{$\alpha$}                                                                                    &   \multicolumn{1}{c}{-1.55}          &   \multicolumn{1}{c}{-1.28}              &     \multicolumn{1}{c}{-1.44}        \\
                                                                                                                                                                                                                                                                   \\
\multicolumn{1}{c}{(4)} & \multicolumn{1}{c}{$M^{*}-5\log h$}                                                                             &   \multicolumn{1}{c}{-20.49}         &   \multicolumn{1}{c}{-20.70}             &     \multicolumn{1}{c}{-21.63}       \\
                                                                                                                                                                                                                                                                   \\
\multicolumn{1}{c}{(5)} & \multicolumn{1}{c}{$M/L$ [M$_{\odot}$/L$_{\odot}]$}                                                             &   \multicolumn{1}{c}{479}            &   \multicolumn{1}{c}{394}                &     \multicolumn{1}{c}{328}          \\
                                                                                                                                                                                                                                                                   \\
\multicolumn{1}{c}{(6)} & \multicolumn{1}{c}{${L_{{\rm{BG1}}}}/{L_{{\rm{tot}}}}$}                                                         &   \multicolumn{1}{c}{0.26 (0.36)}    &   \multicolumn{1}{c}{0.29 (0.37)}        &     \multicolumn{1}{c}{0.30 (0.38)}  \\
\hline            
\end{tabular*}     
\begin{footnotesize}
\begin{flushleft}
{\bf  Notes:}  Luminosities are given in $10^{11}$L$_{\odot}$. \\
(1):  Total luminosity of all member galaxies (hatched histograms). \\
(2):  Total luminosity derived from the best-fit Schechter function. \\
(3) and (4):  Schechter function parameters. \\
(5):  Mass-to-light ratios are calculated with the mean group mass and the luminosities from (1). \\
(6):  Luminosity fraction of the central elliptical compared to the overall group luminosity. Numbers in parantheses refer to spectroscopically confirmed members only.   \\
\end{flushleft}
\end{footnotesize}
\end{minipage}
\end              {table}

\subsection{Mass-to-light ratio}
Total     group     luminosities   have        been     estimated  by     adding     up  the    luminosities     of   spectroscopically   confirmed  group  members   and    galaxies    photometrically
classified  as    members.   In addition, group    luminosities  have    also    been   determined     by      integrating       the       best-fit       Schechter       function         over      all
luminosity bins.   The luminosity of  the central elliptical was  added  to these latter values since  the galaxy was  previously excluded  for the fit. Luminosities have  been derived   assuming that
the   absolute    magnitude      of    the    sun     is     $M_{g^{\prime}\odot}=5.14$,    $M_{r^{\prime}\odot}=4.65$,      and     $M_{i^{\prime}\odot}=4.54$\footnote{Calculated     by           C.\
Willmer:\\   \url  {http://mips.as.arizona.edu/~cnaw/sun.html}}.     Galaxy   magnitudes     were  corrected      for  Galactic     extinction  ($A_{g^{\prime}}=0.15$,     $A_{r^{\prime}}=0.11$,   and
$A_{i^{\prime}}=0.08$) and $k$-correction  ($k_{g^{\prime}}=0.22$,   $k_{r^{\prime}}=0.09$,  and   $k_{i^{\prime}}=0.04$)  before  converting  them   to   luminosities\footnote{Correction  values  are
averages determined   from  the  corresponding SDSS   entries  of  all  spectroscopically  confirmed  group   members.}. The   luminosity fraction  of     the   central  elliptical  compared    to the
overall   group    luminosity  budget     amounts   to     $\sim$30\%   which     is  significantly   lower  than   what  has    been  observed    for  the    fossil  RX~J1340.6+4018      ($\sim$80\%;
\citealt{mendesdeoliveira09}),  most     likely   due    to   the    presence  of     BG2 and  the steep   luminosity function. To estimate   the  mass-to-light ratio  of the
system,   the  mean  group  mass  listed in  Table  \ref{lumdynamics}    was  considered.    All luminosity   estimates    and  the   resulting      mass-to-light  ratios  in   the SDSS  $g^{\prime}$,
$r^{\prime}$, and $i^{\prime}$  bands  are   shown in   Table   \ref{masstolight}.  The  luminosity    fraction of   the central  elliptical  compared    to  the  overall   group   luminosity   budget
is  also  presented.     The      resulting   mass-to-light
ratios     are   comparatively high showing   values around $M/L\sim 400$  M$_{\odot}/$L$_{\odot}$  similar to the findings from other fossils (\citealt{vikhlinin,khosroshahi07})\footnote{The  derived
$r'$ band $M/L$ ratio    was converted to a   corresponding   value  in $B$ to  compare the result with the $M/L$ ratios from \cite{khosroshahi07}. Uniform  colours   of  $(B-R)=1.5$ (assuming
all  satellites to   be   early-types)   and $(B -R)=0.8$  (assuming  all satellites  to be    late-types) were adopted,   yielding     upper and lower limits in the    resulting      $B$-band $M/L$
 ratio of $200< M/L_{B}$ [M$_{\odot}/$L$_{\odot}] <600$.}.

\subsection{Scaling relations}

\begin {figure*}[t]
\begin {center}
\includegraphics[width=\textwidth]{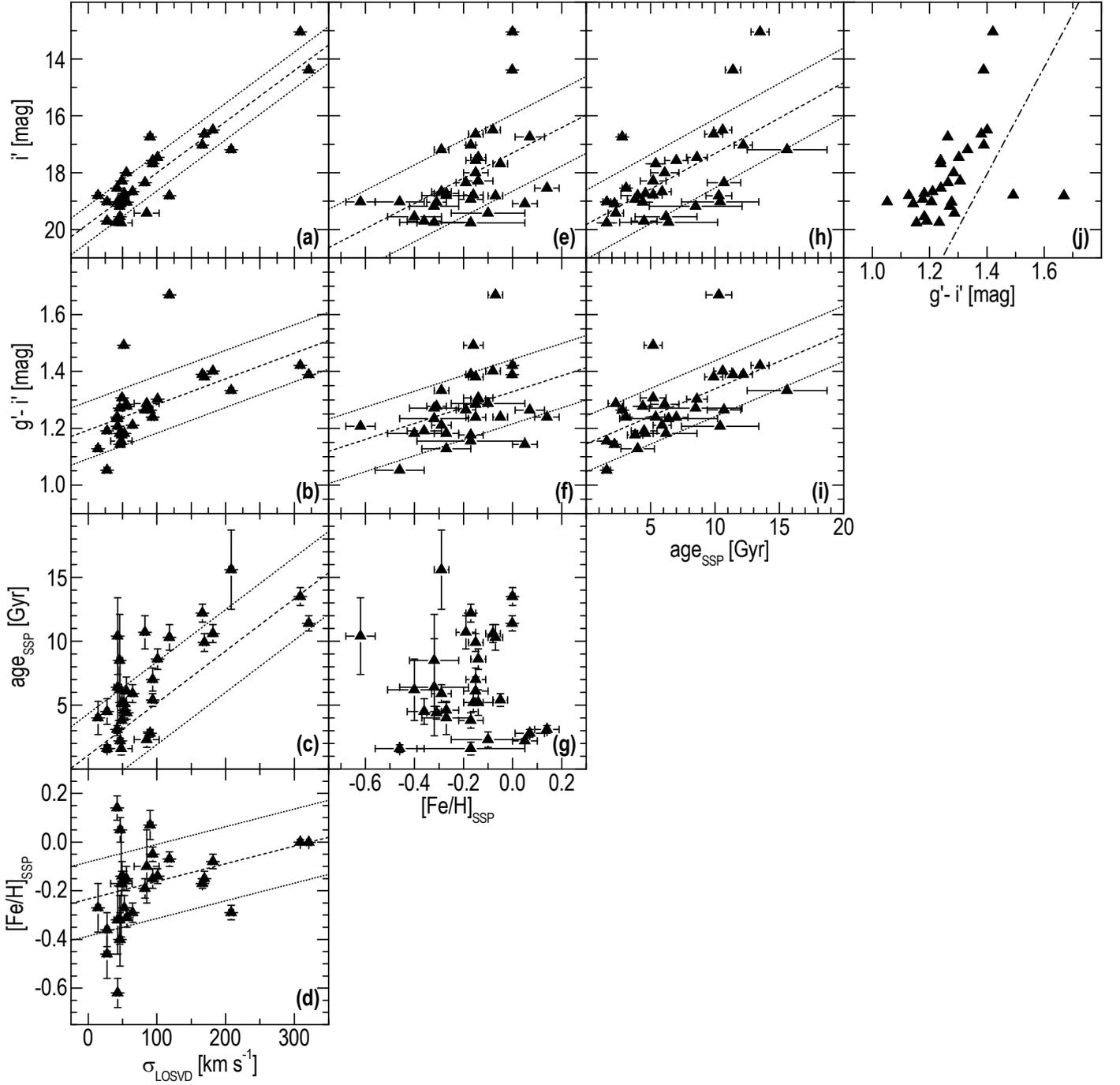}
\caption{\label{dwarfrelations}  Scaling relations  for the   studied galaxy  population in the  RX~J1548.9+0851 system. Dashed lines are  linear
least squares fits to the data. Dotted lines indicate one-sigma confidence intervals. The dash-dotted line in panel {\bf(j)} is the red sequence upper limit as shown in Fig.\ \ref{cmds}.}
\end {center}
\end {figure*}

Figure    \ref{dwarfrelations}   shows    scaling  relations     between  galaxy  brightness,   colour,  line-of-sight  velocity   dispersion,   SSP   age    and  metallicity  for   the   investigated
RX~J1548.9+0851 galaxy  population.   Velocity  dispersions   have been   derived  with the spectrum-fitting packge  ULySS,  providing   the     amount of    broadening applied   to  an  SSP model  to
match  a galaxy spectrum.  Physically meaningful dispersions have  been estimated by  correcting   this broadening for the  instrumental dispersion of the  spectrograph and the dispersion  of  the SSP
model. Panels {\bf  a)}$-${\bf  d)} relate  the  measured    galaxy  velocity  dispersions  with    the remaining   parameters.  The diagrams  show  that galaxies   tend  to     be brighter,   redder,
older, and   more metal   rich with increasing velocity  dispersion. Panels  {\bf  e)}    and  {\bf  f)}  indicate     a  slight    increase    of  galaxy brightness and  colour   with    metallicity.
No  trend    is observed   when    correlating    SSP ages    with     SSP  metallicities.   Panels  {\bf   h)}  and    {\bf   i)}    show  a   clear   trend between brightness, colour,  and      age.
Especially  colour shows a  comparatively   tight   relation with  age, only  members     39  and 42  in   the   vicinity    of the   two  brightest    ellipticals being systematically redder.   Panel
{\bf    j)}   shows the  system's  red   sequence,   the    dash-dotted   line   presenting the   adopted red   sequence   upper   limit  as   seen in  Fig.\  \ref{cmds}.    The presented      scaling
relations      clearly  highlight  the  fact  that     more massive     galaxies (higher  velocity   dispersions),    are     redder,      brighter,  older     and   more
metal   rich compared  to less  massive    ones,  implying   also the  observed relations   without velocity     dispersion    (mass).   This is     plausible  since     more-massive   galaxies,  i.e.
ellipticals,  exhibit  higher luminosities when   assuming  a  constant mass-to-light  ratio,  can   retain   more  metals  due   to  their  deeper  potential  wells,    and  are  passively   evolving
with a  dominating   old stellar population absent of any  young stars dominating the galaxy light in the blue.

\section{Discussion}
\begin {figure}
\begin {center}
\includegraphics[width=\columnwidth]{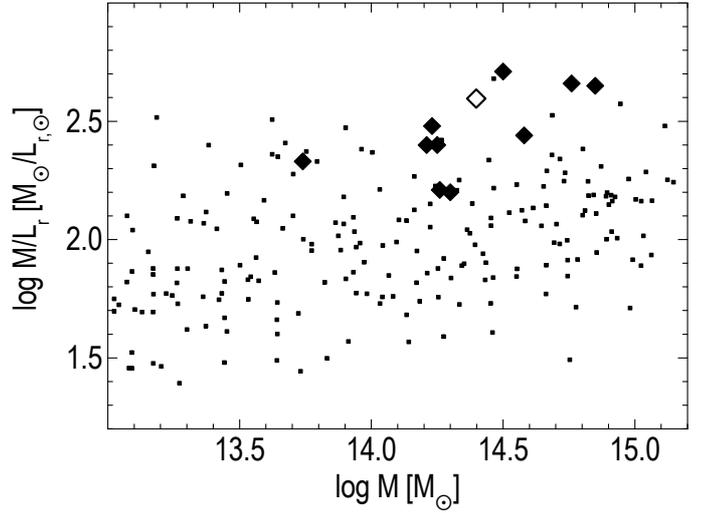}
\caption{\label{mlplot}   Mass-to-light  ratios  plotted  against  dynamical  mass. Black  squares  are  groups  and  clusters  from \citet{girardi02}  while  black diamonds are  the  fossils from
\citet{proctor11}. RX~J1548.9+0851 is shown as open diamond.}
\end {center}
\end {figure}

Using VIMOS  multi-object spectroscopy,  the galaxy   population of  the   fossil  candidate RX~J1548.9+0851   has been  studied in  this work.     The system was selected from the   sample of  fossil
candidates presented  in   \citet{santos}.  At a   redshift  of   $z=0.0721$,  the   system  lies   at a      luminosity distance   of   $D_{L}=326$   Mpc  and    an angular   diameter  distance    of
$D_{A}=283$  Mpc  corresponding to  a  spatial  scale of  1.374 kpc  arcsec$^{-1}$.  At  this redshift  the age  of the  universe amounts  to $\sim   12.5$ Gyr.\\  The      optical     appearance   of
RX~J1548.9+0851      is      dominated   by      two    giant elliptical       galaxies  in the central  region  of the system, exhibiting        a  magnitude difference of $\Delta m_{1,2}=1.34$    in
the SDSS $r'$    band  and   a  spectroscopic   redshift  difference    of  $\Delta  z   \sim   0.004$ (SDSS     DR5). Although this  magnitude  difference   at   a projected distance   of    326  kpc
clearly    violates  the  2   mag criterion  for fossils, the  system    is  listed  in  the  sample   of  fossil candidates by \citet{santos}  who use a  constant value   of $0.5   h_{70}^{-1}$   Mpc
to estimate half  the  group virial   radius. The classification of RX~J1548.9+0851   as fossil     in  \citet{santos} is  plausible though  since  these   authors consider     all  satellites    with
$\Delta    z>0.002$       as  nonmember   galaxies.  Following    the      distribution   of     galaxies     in   radial    velocity    space  as     seen    in  Fig.\  \ref{velocityhistogram},   and
considering    the applied  membership criterion   of \citet{ramella94}  ($\Delta    z<0.005$),   there  is  no   reason to classify  the second-brightest elliptical  as background   object   in  this
work,  however.    At    the     group  redshift    of   $z=0.0721$,      the  \citet{evrard96}    estimate   for     half    the       virial  radius      yields    a   projected  distance    of   at
least  331 kpc\footnote{$r_{{\rm{vir}}}   = 1.945  \cdot  \left(  T/10\,\rm{keV} \right)^{1/2}  \left( {1  +  z}  \right)^{ -  3/2}  \cdot  h^{  -   1} {\rm{  Mpc}}$; Since  no  X-ray  temperature was
derived for RX~J1548.9+0851,  a lower   limit  of   0.7  keV  was  assumed.  This value   is based   on the      gas temperature $-$   bolometric  X-ray  luminosity relation from   the  Millennium gas
simulation  (see Fig.\ 1 in \citealt{dariush07})  showing that  a  value    of 0.7  keV corresponds  to   about  $L_{X,{\rm bol}}=    10^{42}  h^{  -2}$  erg  s$^{-1}$, the  X-ray limit    of  fossils
as proposed    by \citet{jones03}.   This  temperature  limit   seems   reasonable  since     the    lowest   gas temperature     of  the    15    fossils      listed  by    \citet{mendesdeoliveira06}
shows    a  similar    value     (NGC   6482;  \citealt{khosroshahi04}).} which  would   classify  RX~J1548.9+0851   as  non-fossil. Estimating  the virial radius via the luminosity-weighted dynamical
formulae as shown in  Table \ref{lumdynamics} yields a  value  of \nicefrac{1}{2}~$r_{\rm{V}}=305$~kpc, however,  classifying RX~J1548.9+0851   as fossil.  In any    case the  group is    of   special
interest since   it  is lacking any   other  bright galaxies that  would be expected  in  a poor  group environment. Multi-object spectroscopy of the RX~J1548.9+0851 galaxy population has   revealed a
luminosity-weighted velocity dispersion of 568 km  s$^{-1}$ and  a mass of $\sim 2.5 \times 10^{14}$ M$_{\odot}$   for the   system.    This    result    confirms   previous    studies    that fossils
seem    to    be     very   massive  (\citealt{mendesdeoliveira06,mendesdeoliveira09,cypriano,khosroshahi07,proctor11}) classifying them  as fossil clusters  rather than groups  inconsistent with  the
proposed  merging   scenario.       Figure   \ref{mlplot}  shows  mass-to-light  ratios   and  dynamical   masses  of   ordinary  poor   groups  and   clusters  from \citet{girardi02} and   10 fossils
from  \citet{proctor11}. Fossils  occupy the   upper envelope  of  the  distribution which  had already   been previously   discovered   by  \citet{khosroshahi07}. The   mass-to-light      ratio   of
RX~J1548.9+0851 confirms these  studies, showing a value of     $M/L_{r}\sim400$M$_{\odot}/$L$_{r,\odot}$.

\begin {figure*}
\begin {center}
\includegraphics[width=400pt]{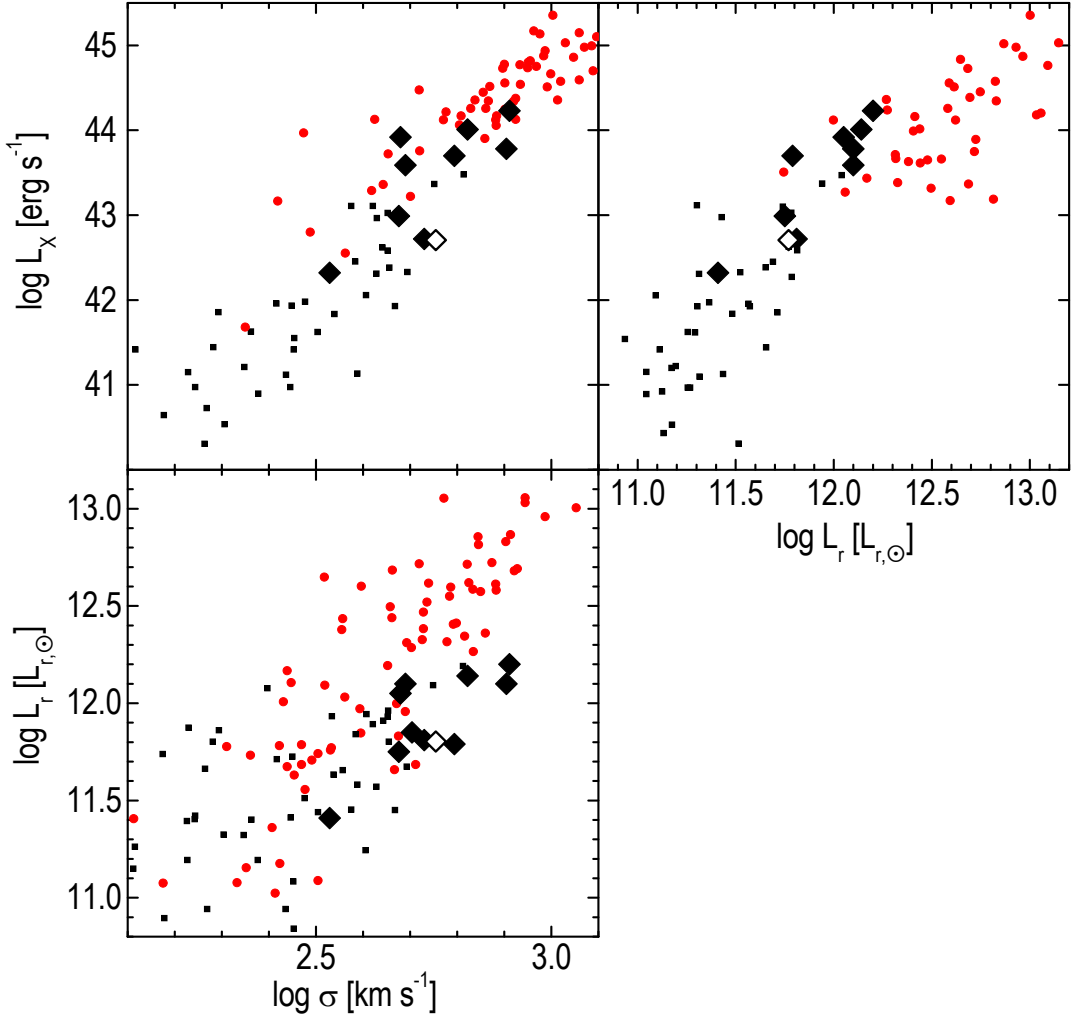}
\caption{\label{scalinggroup}  The RX~J1548.9+0851 system in scaling relations  of ordinary groups and clusters. Black squares  are groups from \citet{osmond} and \citet{girardi02} while  red dots are
clusters from \citet{girardi02} and \citet{wu}. Black diamonds are fossils from \citet{proctor11}.  RX~J1548.9+0851 is shown as open diamond.}
\end {center}
\end {figure*}

\citet{khosroshahi07}  and \citet{proctor11}  have studied  scaling relations  of fossils  relating  X-ray  luminosity $L_{X}$,  optical luminosity  $L_{r}$, and  velocity dispersion  $\sigma$.  While
\citet{khosroshahi07} found that,  for a given  velocity dispersion, fossils  are overluminous in  X-rays, but comparable  to ordinary groups  in $L_{r}$, \citet{proctor11}  claimed that for  a  given
velocity dispersion, fossils  are  underluminous  in  the optical,  but otherwise  fall on  the $L_{X}-\sigma$ relation  of  groups and clusters. These results led to contrary interpretations,  either
fossils formed early, building  up  a  cuspy dark  matter halo,  leading to   the excess in  X-ray  luminosity (\citealt{khosroshahi07}),  or  they  are cluster-like  in  their masses,  but  otherwise
possess the richness  and optical luminosity  of relatively poor  groups (\citealt{proctor11}). Figure  \ref{scalinggroup} shows  the  location of  the  RX~J1548.9+0851 system  in  the $L_{X}-\sigma$,
$L_{r}-\sigma$, and $L_{X}-L_{r}$  planes. In  all     diagrams,     RX~J1548.9+0851     is  found within  the    distribution  of  fossils from \citet{proctor11},  being also clearly underluminous in
the optical.  Only in the $L_{X}-\sigma$  plane,     RX~J1548.9+0851   is   not as X-ray    luminous with  respect    to   other fossils.  However, X-ray  luminosities  are  derived  from   low-signal
ROSAT     count rates  which  aren't  the best  quality data   to  accurately  estimate  the   overall  X-ray  luminosity. The  present findings  on RX~J1548.9+0851  support  the  interpretation  from
\citet{proctor11}, leaving the question of what happened to all the missing baryons.

The colour-magnitude diagram of RX~J1548.9+0851 reveals  a pronounced red  sequence and a  lack of blue galaxies within the   system. In fact, 90\% of all  spectroscopically confirmed member  galaxies
in the inner  300 kpc  of the group are found  within a narrow band of  $\sim 0.3$ mag  thickness in  colour. Galaxies bluer than this  distribution are at least 6  magnitudes fainter than the central
elliptical. Moving out to 1 Mpc,  still 89\% of all spectroscopically  confirmed members are found within  this stripe, slightly increasing in  thickness to 0.4 mag. Again,  blue galaxies outside this
region  are $\sim 6$ magnitudes  fainter than the  central  galaxy and  constitute of only  4 objects (see  Fig.\  \ref{cmds}).  This  dominance of red  galaxies within the  group  suggests  that
most of these galaxies, especially the  luminous ones, are  old and passively  evolving.  A similar  result was  found  in the CMDs of the two  fossil groups CXGG  095951+0140.8 and  CXGG095951+0212.6
at  $z\sim0.4$ discovered in the  Cosmic Evolution Survey  (COSMOS) by  \citet{pierini11}. Only     2 members     are  found   above   the  adopted red  sequence  upper limit. Interestingly,  these  2
galaxies   are located in   the immediate vicinity  of the  two brightest ellipticals  and show a compact, circular appearance.   A similar object has been found around  the central  elliptical in the
evolved  NGC 5846 group within  the Local Supercluster (\citealt{eigenthaler10}) suggesting that these satellites have been tidally stripped by their massive host galaxies.

Symmetric   shells are  revealed  unambiguously along  the major  axis  of   the central  elliptical.    The   presence  of  these   symmetric   shells  indicates     a   recent  minor merger  in the
center  of the system.   In   fact,  \citet{ebrova} have simulated   shell   galaxies with   the conclusion that  a   significantly smaller   secondary   galaxy interacting    with  the dark   matter
halo  of  the  primary  elliptical  can reproduce such observed regular shell    systems.  In their  simulation, a  satellite  falling  radially   towards  the  center of   the primary  galaxy     is
assumed  to be instantly    ripped  apart when  approaching the   center  of   the   primary.  Its stars   subesequently   oscillate  in   the  potential    of the  primary elliptical    and  produce
arc-like  structures   at  the    orbit turning   points  as observed.  Thus,   being  the    most massive  object   in  the  RX~J1548.9+0851 system  and   exhibiting  the observed  fine   structure,
it is   very  likely   that  a     recent minor   merger   occured  in  this   system. This   would be  in agreement    with the   evolution  scenario for  fossils   as  proposed by \citet{dariush07}
suggesting   that   these aggregates     have assembled   most  of    their mass     at early epochs  and   grow typically      by minor  mergers afterwards. 

Existing   studies have    revealed  steep,    increasing    luminosity   functions for   the fossil  clusters     RX~J1340.6+4018  ($\alpha=    -1.6    \pm    0.2$; \citealt{mendesdeoliveira09})  and
RX~J1416.4+2315 ($\alpha=  -1.2$ to  $-1.3$; \citealt{cypriano}),  consistent with    those of  other clusters    (e.g.\ Virgo  and Coma).    On the  other hand,   a decreasing  faint-end slope    has
been found for    the fossil  RX~J1552.2+2013 ($\alpha=  -0.4$   to  $-0.8$; \citealt{mendesdeoliveira06}). Interestingly,  this   fossil  is also   the only  one  in the literature to  exhibit  a  cD
envelope around the  central elliptical  and a  dip  in  the luminosity  function around  $M_{r^{\prime}}=-18$. Hence,  \citet{mendesdeoliveira06}  argued  that the  presence of  a cD envelope  around
the central elliptical could be related to  the lack of faint galaxies and the  steepness of the faint-end luminosity function. The RX~J1548.9+0851 system  studied in  this work also revealed a steep,
increasing luminosity function,  exhibiting   high values  around  $\alpha=  -1.4  \pm    0.1$.  No   lack of faint galaxies,  i.e.\  a   pronounced dip  around  $M_{r^{\prime}}=-18 + 5\log   h$   has
been found.  The surface  brightness   profile  of  the central elliptical    is well-fit  by  a pure  deVaucouleurs   $r^{1/4}$ law  without  any    indication of   a  light   excess    in  the outer
galaxy regions, consistent with the suggested scenario highlighted in the previous studies.

Spectra of RX~J1548.9+0851  member galaxies  have  been fit with  SSP   models to  determine SSP  equivalent   ages and metallicities  of the   group galaxy population. The  distribution  of  SSP ages
within the system  shows a clear   spatial  segregation.  While galaxies exhibiting  an   old ($>  8$   Gyr) stellar population   are  confined to an  elongated,  central  distribution, galaxies  with
young  ($< 8$  Gyr)  stars  are   more  diffusely  distributed   and  predominantly  found   in  the  outer   regions  of  the   system.  Interestingly, the    distribution   of  old  galaxies is also
correlated  with  the orientation   of  the major  axis of the central elliptical.  This  is in  agreement with   the work  of \citet{sales07}  who used   the  Millennium  simulation  to   investigate
correlations   between  the  central    \emph{primary}   galaxies  and   the   surrounding  galaxy populations. They   found that   red  satellites   are  clearly much more    concentrated towards   a
massive central   galaxy  than    blue ones.    They argue   that  this  is  presumably    due   to  the   gas   loss  of  satellites   once    they   are  accreted.   The earlier  a   satellite   was
accreted,   the    older,     i.e.\     redder,   its    stellar   population   will   be.    This    suggests     that     the  distribution   of   old   stellar   population  galaxies      in    the
RX~J1548.9+0851  system    was formed    first, likely   representing   a  virialized   substructure whereas  galaxies with   younger  stellar  populations are  not in  dynamical equilibrium    within
the group potential.   The  simulation   also revealed      a systematic anisotropy  in    the  spatial     distribution  of satellites  around primaries.     It     has     been  established     that
dark   matter    haloes     are   triaxial   objects,  preferentially exhibiting     prolate        shapes whose     angular     momentum   is   perpendicular     to  the     major-axis       of   the
halo     (\citealt{hopkins}).   Simulated    satellites    show   an anti-Holmberg  effect, being aligned      along the    plane perpendicular    to  the angular momentum   axis of   the halo.  Thus,
based  on this simulation and  assuming  that the major  axis of a primary  is  aligned  with   the   major  axis of   the  dark matter   halo,  an  alignment  of   satellites with    the  major  axis
of  the  central   galaxy  in fossils  is  expected. The   results presented in Fig.\ref{agemetallicitymaps} show such an alignment. The measured  SSP ages     are  also correlated with   the observed
radial velocity distribution.      Old galaxies are  mainly found at  velocities around  the   two  brightest ellipticals  BG1    and BG2  which are   also  the  oldest objects   in the  system.    In
contrast,  young objects show  a smooth distribution  over     all   velocities and  exhibit   a     much  larger  dispersion  supporting the highlighted scenario where   the old  galaxy    population
assembled  first,  leaving enough   time    to  virialize  around  the  most massive   subhaloes BG1  and  BG2.

\section{Conclusion}
We have presented  a spectroscopic analysis  of the fossil  candidate RX~J1548.9+0851, selected  from  the sample  of \citet{santos}. The  system is dominated  by two bright ellipticals in the center,
questioning the definition as  fossil. Nonetheless, similarities to  other fossils studied recently  have been found, given  that the system is  rather comparable to clusters  in mass,  shows a  steep
faint-end luminosity function,  high mass-to-light ratio,  and a  central  elliptical without cD  envelope. Ages and  radial  velocities of member galaxies suggest that satellites hosting  old stellar
populations form a virialized substructure. Shells  around the central elliptical indicate the  presence of a recent minor merger.  Comparing RX~J1548.9+0851 with scaling relations from  ordinary poor
groups and clusters confirm the idea  that fossil might simply be normal  clusters with the richness  of poor  groups. The missing baryons could have  been expelled from the system or are   hidden  in
the intracluster light.

\begin{acknowledgements}
We acknowledge the useful comments from the anonymous referee which helped to improve the paper. We are also very grateful to the language editor, xxxx, for his/her careful reading of the  manuscript.
P.\ E.\ was supported by the University of Vienna in the frame of the Initiative Kolleg (IK) \textit{The Cosmic Matter Circuit} I033-N. This work is based on observations made with the ESO-VLT at  the
La Silla Paranal Observatory under programme ID 383.A-0123 and  made use of the astronomical data reduction software IRAF which is distributed by the National Optical Astronomy Observatories (NOAO).
\end{acknowledgements}

\begin{table*}
\begin{minipage}[t]{520pt}
\caption{\label{members}Identified group  members  in   the observed VIMOS  field within 1 Mpc of the central elliptical.}
\centering          
\renewcommand{\footnoterule}{}         
\begin            {tabular*}{520pt}{@{\extracolsep{\fill}}p{1.5cm}p{1.5cm}p{1.4cm}p{0.6cm}p{0.6cm}p{1.4cm}p{1cm}p{1.5cm}p{1.5cm}p{2cm}}
\hline          
\multicolumn{1}{c}{\multirow{2}{*}{galaxy$^{a}$}}       &   \multicolumn{1}{c}{\multirow{2}{*}{${\alpha _{{\rm{J2000}}}}^b$}}    &   \multicolumn{1}{c}{\multirow{2}{*}{${\delta _{{\rm{J2000}}}}^b$}}   &    \multicolumn{1}{c}{$i'^{c}$}     &  \multicolumn{1}{c}{$g'-i'^{c}$}    &         \multicolumn{1}{c}{${v_{{\rm{rad}}}}^d$}        &    \multicolumn{1}{c}{\multirow{2}{*}{$R^{d}$}}   &            \multicolumn{1}{c}{${\rm{age}_{{\rm{SSP}}}}^e$} &     \multicolumn{1}{c}{${\rm{[Fe/H]}_{{\rm{SSP}}}}^e$}    &    \multicolumn{1}{c}{$\sigma^{e}$}       \\
                                                        &                                                                        &                                                                       &   \multicolumn{1}{c}{[mag]}         &    \multicolumn{1}{c}{[mag]}        &         \multicolumn{1}{c}{[km s$^{-1}$]}               &                                                   &          \multicolumn{1}{c}{[Gyr]}                         &        \multicolumn{1}{c}{dex}                            &  \multicolumn{1}{c}{[km s$^{-1}$]}        \\
\hline                                                                                                                                                                                                                                                                                                                                                                                                                                                                                                                                                                    
01\dotfill                                              &      \centering{15 48 42.24}                                           &    \centering{08 52 26.3}                                             &        \centering{18.81}            &          \centering{1.13}           &        \centering{19899$\,\pm\,$26}                     &  \multicolumn{1}{r}{ 5.70}                        &  \multicolumn{1}{r}{ 4.0$\,\pm\,$1.3}                      &     \multicolumn{1}{r}{$-$0.27$\,\pm\,$0.10}              &  \multicolumn{1}{c}{14}                   \\
03\dotfill                                              &      \centering{15 48 43.86}                                           &    \centering{08 48 42.5}                                             &        \centering{19.18}            &          \centering{1.27}           &        \centering{20960$\,\pm\,$27}                     &  \multicolumn{1}{r}{ 4.25}                        &  \multicolumn{1}{r}{ 8.5$\,\pm\,$3.6}                      &     \multicolumn{1}{r}{$-$0.32$\,\pm\,$0.10}              &  \multicolumn{1}{c}{46}                   \\
07\dotfill                                              &      \centering{15 48 45.70}                                           &    \centering{08 49 03.8}                                             &        \centering{19.75}            &          \centering{1.23}           &        \centering{21172$\,\pm\,$33}                     &  \multicolumn{1}{r}{ 4.28}                        &  \multicolumn{1}{r}{ 6.4$\,\pm\,$3.8}                      &     \multicolumn{1}{r}{$-$0.32$\,\pm\,$0.14}              &  \multicolumn{1}{c}{43}                   \\
09\dotfill                                              &      \centering{15 48 45.99}                                           &    \centering{08 43 24.6}                                             &        \centering{16.94}            &          \centering{1.36}           &        \centering{21183$\,\pm\,$27}                     &  \multicolumn{1}{r}{ 5.64}                        &  \multicolumn{1}{r}{ 3.6$\,\pm\,$0.7}                      &     \multicolumn{1}{r}{$-$0.07$\,\pm\,$0.05}              &  \multicolumn{1}{c}{108}                  \\
12\dotfill                                              &      \centering{15 48 49.20}                                           &    \centering{08 48 53.0}                                             &        \centering{19.03}            &          \centering{1.05}           &        \centering{19469$\,\pm\,$28}                     &  \multicolumn{1}{r}{ 4.04}                        &  \multicolumn{1}{r}{ 1.6$\,\pm\,$0.3}                      &     \multicolumn{1}{r}{$-$0.46$\,\pm\,$0.10}              &  \multicolumn{1}{c}{ 27}                  \\
14\dotfill                                              &      \centering{15 48 49.51}                                           &    \centering{08 54 21.9}                                             &        \centering{16.75}            &          \centering{1.26}           &        \centering{21716$\,\pm\,$42}                     &  \multicolumn{1}{r}{ 5.65}                        &  \multicolumn{1}{r}{ 2.8$\,\pm\,$0.3}                      &     \multicolumn{1}{r}{0.07$\,\pm\,$0.06}                 &  \multicolumn{1}{c}{ 90}                  \\
17\dotfill                                              &      \centering{15 48 50.37}                                           &    \centering{08 49 12.6}                                             &        \centering{18.00}            &          \centering{1.28}           &        \centering{21512$\,\pm\,$23}                     &  \multicolumn{1}{r}{ 5.36}                        &  \multicolumn{1}{r}{ 6.1$\,\pm\,$1.1}                      &     \multicolumn{1}{r}{$-$0.15$\,\pm\,$0.05}              &  \multicolumn{1}{c}{ 56}                  \\
19\dotfill                                              &      \centering{15 48 50.55}                                           &    \centering{08 48 28.9}                                             &        \centering{17.19}            &          \centering{1.33}           &        \centering{20599$\,\pm\,$32}                     &  \multicolumn{1}{r}{ 5.73}                        &  \multicolumn{1}{r}{15.6$\,\pm\,$3.1}                      &     \multicolumn{1}{r}{$-$0.29$\,\pm\,$0.03}              &  \multicolumn{1}{c}{208}                  \\
20\dotfill                                              &      \centering{15 48 50.73}                                           &    \centering{08 42 48.4}                                             &        \centering{17.89}            &          \centering{1.30}           &        \centering{20632$\,\pm\,$36}                     &  \multicolumn{1}{r}{ 5.72}                        &  \multicolumn{1}{r}{ 4.7$\,\pm\,$1.5}                      &     \multicolumn{1}{r}{$-$0.26$\,\pm\,$0.07}              &  \multicolumn{1}{c}{ 47}                  \\
22\dotfill                                              &      \centering{15 48 51.39}                                           &    \centering{08 53 04.0}                                             &        \centering{19.02}            &          \centering{1.21}           &        \centering{19889$\,\pm\,$23}                     &  \multicolumn{1}{r}{ 5.05}                        &  \multicolumn{1}{r}{10.4$\,\pm\,$3.0}                      &     \multicolumn{1}{r}{$-$0.62$\,\pm\,$0.06}              &  \multicolumn{1}{c}{ 43}                  \\
24\dotfill                                              &      \centering{15 48 53.74}                                           &    \centering{08 51 37.7}                                             &        \centering{16.64}            &          \centering{1.38}           &        \centering{20983$\,\pm\,$24}                     &  \multicolumn{1}{r}{ 9.34}                        &  \multicolumn{1}{r}{ 9.9$\,\pm\,$0.7}                      &     \multicolumn{1}{r}{$-$0.15$\,\pm\,$0.03}              &  \multicolumn{1}{c}{169}                  \\
25\dotfill                                              &      \centering{15 48 53.78}                                           &    \centering{08 51 51.5}                                             &        \centering{17.57}            &          \centering{1.24}           &        \centering{19572$\,\pm\,$26}                     &  \multicolumn{1}{r}{10.07}                        &  \multicolumn{1}{r}{ 7.0$\,\pm\,$0.9}                      &     \multicolumn{1}{r}{$-$0.15$\,\pm\,$0.04}              &  \multicolumn{1}{c}{ 94}                  \\
26\dotfill                                              &      \centering{15 48 54.75}                                           &    \centering{08 53 08.2}                                             &        \centering{19.09}            &          \centering{1.14}           &        \centering{21851$\,\pm\,$74}                     &  \multicolumn{1}{r}{ 6.65}                        &  \multicolumn{1}{r}{ 2.2$\,\pm\,$0.2}                      &     \multicolumn{1}{r}{0.05$\,\pm\,$0.05}                 &  \multicolumn{1}{c}{ 46}                  \\
27\dotfill                                              &      \centering{15 48 54.77}                                           &    \centering{08 52 57.5}                                             &        \centering{18.35}            &          \centering{1.26}           &        \centering{20793$\,\pm\,$26}                     &  \multicolumn{1}{r}{10.65}                        &  \multicolumn{1}{r}{10.7$\,\pm\,$1.3}                      &     \multicolumn{1}{r}{$-$0.19$\,\pm\,$0.04}              &  \multicolumn{1}{c}{ 83}                  \\
28\dotfill                                              &      \centering{15 48 54.85}                                           &    \centering{08 49 02.9}                                             &        \centering{19.55}            &          \centering{1.18}           &        \centering{21776$\,\pm\,$96}                     &  \multicolumn{1}{r}{ 6.30}                        &  \multicolumn{1}{r}{ 6.2$\,\pm\,$2.4}                      &     \multicolumn{1}{r}{$-$0.40$\,\pm\,$0.11}              &  \multicolumn{1}{c}{ 46}                  \\
29\dotfill                                              &      \centering{15 48 54.93}                                           &    \centering{08 49 43.0}                                             &        \centering{17.03}            &          \centering{1.39}           &        \centering{20648$\,\pm\,$30}                     &  \multicolumn{1}{r}{10.11}                        &  \multicolumn{1}{r}{12.2$\,\pm\,$0.7}                      &     \multicolumn{1}{r}{$-$0.17$\,\pm\,$0.02}              &  \multicolumn{1}{c}{166}                  \\
30\dotfill                                              &      \centering{15 48 55.12}                                           &    \centering{08 48 03.5}                                             &        \centering{18.94}            &          \centering{1.18}           &        \centering{21049$\,\pm\,$24}                     &  \multicolumn{1}{r}{ 8.61}                        &  \multicolumn{1}{r}{ 3.8$\,\pm\,$0.6}                      &     \multicolumn{1}{r}{$-$0.17$\,\pm\,$0.05}              &  \multicolumn{1}{c}{ 49}                  \\
31\dotfill                                              &      \centering{15 48 55.44}                                           &    \centering{08 54 08.9}                                             &        \centering{19.04}            &          \centering{1.28}           &        \centering{20303$\,\pm\,$29}                     &  \multicolumn{1}{r}{ 7.99}                        &  \multicolumn{1}{r}{ 4.4$\,\pm\,$0.5}                      &     \multicolumn{1}{r}{$-$0.31$\,\pm\,$0.04}              &  \multicolumn{1}{c}{ 56}                  \\
32\dotfill                                              &      \centering{15 48 55.82}                                           &    \centering{08 49 10.6}                                             &        \centering{16.50}            &          \centering{1.40}           &        \centering{20205$\,\pm\,$27}                     &  \multicolumn{1}{r}{11.39}                        &  \multicolumn{1}{r}{10.6$\,\pm\,$0.7}                      &     \multicolumn{1}{r}{$-$0.08$\,\pm\,$0.03}              &  \multicolumn{1}{c}{182}                  \\
34\dotfill BG1\dotfill                                  &      \centering{15 48 55.96}                                           &    \centering{08 50 42.9}                                             &        \centering{13.05}            &          \centering{1.42}           &        \centering{20854$\,\pm\,$31}                     &  \multicolumn{1}{r}{10.15}                        &  \multicolumn{1}{r}{13.5$\,\pm\,$0.7}                      &     \multicolumn{1}{r}{0.00$\,\pm\,$0.01}                 &  \multicolumn{1}{c}{309}                  \\
35\dotfill                                              &      \centering{15 48 55.97}                                           &    \centering{08 49 27.2}                                             &        \centering{19.71}            &          \centering{1.19}           &        \centering{20234$\,\pm\,$26}                     &  \multicolumn{1}{r}{ 7.84}                        &  \multicolumn{1}{r}{ 4.5$\,\pm\,$1.0}                      &     \multicolumn{1}{r}{$-$0.36$\,\pm\,$0.07}              &  \multicolumn{1}{c}{ 27}                  \\
36\dotfill                                              &      \centering{15 48 56.71}                                           &    \centering{08 44 58.3}                                             &        \centering{16.68}            &          \centering{1.30}           &        \centering{20558$\,\pm\,$26}                     &  \multicolumn{1}{r}{10.39}                        &  \multicolumn{1}{r}{ 7.2$\,\pm\,$0.9}                      &     \multicolumn{1}{r}{$-$0.09$\,\pm\,$0.04}              &  \multicolumn{1}{c}{119}                  \\
39\dotfill                                              &      \centering{15 48 57.87}                                           &    \centering{08 50 47.3}                                             &        \centering{18.78}            &          \centering{1.49}           &        \centering{20815$\,\pm\,$22}                     &  \multicolumn{1}{r}{10.87}                        &  \multicolumn{1}{r}{ 5.2$\,\pm\,$0.7}                      &     \multicolumn{1}{r}{$-$0.16$\,\pm\,$0.04}              &  \multicolumn{1}{c}{ 52}                  \\
41\dotfill                                              &      \centering{15 48 58.45}                                           &    \centering{08 51 02.4}                                             &        \centering{19.42}            &          \centering{1.29}           &        \centering{19489$\,\pm\,$53}                     &  \multicolumn{1}{r}{ 4.90}                        &  \multicolumn{1}{r}{ 2.3$\,\pm\,$0.6}                      &     \multicolumn{1}{r}{$-$0.10$\,\pm\,$0.15}              &  \multicolumn{1}{c}{ 85}                  \\
42\dotfill                                              &      \centering{15 48 58.80}                                           &    \centering{08 54 27.9}                                             &        \centering{18.81}            &          \centering{1.67}           &        \centering{19464$\,\pm\,$30}                     &  \multicolumn{1}{r}{12.90}                        &  \multicolumn{1}{r}{10.3$\,\pm\,$1.0}                      &     \multicolumn{1}{r}{$-$0.07$\,\pm\,$0.03}              &  \multicolumn{1}{c}{118}                  \\
43\dotfill                                              &      \centering{15 48 58.88}                                           &    \centering{08 52 50.6}                                             &        \centering{17.68}            &          \centering{1.24}           &        \centering{21392$\,\pm\,$28}                     &  \multicolumn{1}{r}{11.88}                        &  \multicolumn{1}{r}{ 5.4$\,\pm\,$0.5}                      &     \multicolumn{1}{r}{$-$0.05$\,\pm\,$0.03}              &  \multicolumn{1}{c}{ 94}                  \\
44\dotfill                                              &      \centering{15 48 59.00}                                           &    \centering{08 52 25.0}                                             &        \centering{19.77}            &          \centering{1.15}           &        \centering{20981$\,\pm\,$32}                     &  \multicolumn{1}{r}{ 5.02}                        &  \multicolumn{1}{r}{ 1.6$\,\pm\,$0.5}                      &     \multicolumn{1}{r}{$-$0.17$\,\pm\,$0.22}              &  \multicolumn{1}{c}{ 48}                  \\
47\dotfill                                              &      \centering{15 48 59.28}                                           &    \centering{08 53 27.5}                                             &        \centering{17.46}            &          \centering{1.30}           &        \centering{20696$\,\pm\,$23}                     &  \multicolumn{1}{r}{12.90}                        &  \multicolumn{1}{r}{ 8.6$\,\pm\,$0.8}                      &     \multicolumn{1}{r}{$-$0.14$\,\pm\,$0.03}              &  \multicolumn{1}{c}{101}                  \\
48\dotfill BG2\dotfill                                  &      \centering{15 48 59.56}                                           &    \centering{08 54 33.2}                                             &        \centering{14.38}            &          \centering{1.39}           &        \centering{19781$\,\pm\,$28}                     &  \multicolumn{1}{r}{10.14}                        &  \multicolumn{1}{r}{11.4$\,\pm\,$0.6}                      &     \multicolumn{1}{r}{$-$0.001$\,\pm\,$0.003}            &  \multicolumn{1}{c}{321}                  \\
51\dotfill                                              &      \centering{15 49 02.01}                                           &    \centering{08 47 56.5}                                             &        \centering{19.57}            &          \centering{0.48}           &        \centering{19606$\,\pm\,$34}                     &  \multicolumn{1}{r}{3.20}                         &  \multicolumn{1}{c}{        $-$     }                      &     \multicolumn{1}{c}{        $-$         }              &  \multicolumn{1}{c}{$-$}                  \\
52\dotfill                                              &      \centering{15 49 02.63}                                           &    \centering{08 42 40.0}                                             &        \centering{19.26}            &          \centering{0.77}           &        \centering{20783$\,\pm\,$42}                     &  \multicolumn{1}{r}{2.92}                         &  \multicolumn{1}{r}{ 3.9$\,\pm\,$3.0}                      &     \multicolumn{1}{r}{$-$0.25$\,\pm\,$0.34}              &  \multicolumn{1}{c}{$-$}                  \\
55\dotfill                                              &      \centering{15 49 04.31}                                           &    \centering{08 54 06.4}                                             &        \centering{18.67}            &          \centering{1.21}           &        \centering{20678$\,\pm\,$25}                     &  \multicolumn{1}{r}{9.90}                         &  \multicolumn{1}{r}{ 5.9$\,\pm\,$0.7}                      &     \multicolumn{1}{r}{$-$0.29$\,\pm\,$0.04}              &  \multicolumn{1}{c}{ 64}                  \\
56\dotfill                                              &      \centering{15 49 05.38}                                           &    \centering{08 53 15.1}                                             &        \centering{18.54}            &          \centering{1.24}           &        \centering{20334$\,\pm\,$31}                     &  \multicolumn{1}{r}{10.55}                        &  \multicolumn{1}{r}{ 3.1$\,\pm\,$0.3}                      &     \multicolumn{1}{r}{ 0.14$\,\pm\,$0.05}                &  \multicolumn{1}{c}{ 42}                  \\
58\dotfill                                              &      \centering{15 49 05.65}                                           &    \centering{08 48 40.1}                                             &        \centering{18.29}            &          \centering{1.31}           &        \centering{21607$\,\pm\,$26}                     &  \multicolumn{1}{r}{11.06}                        &  \multicolumn{1}{r}{ 5.2$\,\pm\,$1.0}                      &     \multicolumn{1}{r}{$-$0.14$\,\pm\,$0.06}              &  \multicolumn{1}{c}{ 49}                  \\
63\dotfill                                              &      \centering{15 49 06.82}                                           &    \centering{08 48 53.2}                                             &        \centering{19.22}            &          \centering{0.50}           &        \centering{21043$\,\pm\,$37}                     &  \multicolumn{1}{r}{3.24}                         &  \multicolumn{1}{c}{        $-$     }                      &     \multicolumn{1}{c}{        $-$         }              &  \multicolumn{1}{c}{$-$}                  \\
64\dotfill                                              &      \centering{15 49 07.05}                                           &    \centering{08 41 37.7}                                             &        \centering{19.13}            &          \centering{0.66}           &        \centering{20225$\,\pm\,$37}                     &  \multicolumn{1}{r}{2.05}                         &  \multicolumn{1}{r}{ 0.7$\,\pm\,$0.1}                      &     \multicolumn{1}{r}{$-$0.59$\,\pm\,$0.12}              &  \multicolumn{1}{c}{$-$}                  \\
67\dotfill                                              &      \centering{15 49 09.47}                                           &    \centering{08 49 19.2}                                             &        \centering{18.73}            &          \centering{1.18}           &        \centering{19801$\,\pm\,$28}                     &  \multicolumn{1}{r}{10.35}                        &  \multicolumn{1}{r}{ 4.6$\,\pm\,$0.6}                      &     \multicolumn{1}{r}{$-$0.27$\,\pm\,$0.05}              &  \multicolumn{1}{c}{ 53}                  \\
71\dotfill                                              &      \centering{15 49 24.05}                                           &    \centering{08 43 44.7}                                             &        \centering{17.06}            &          \centering{1.26}           &        \centering{21512$\,\pm\,$39}                     &  \multicolumn{1}{r}{3.19}                         &  \multicolumn{1}{r}{ 0.8$\,\pm\,$0.2}                      &     \multicolumn{1}{r}{$-$0.31$\,\pm\,$0.15}              &  \multicolumn{1}{c}{ 14}                  \\
76\dotfill                                              &      \centering{15 49 28.82}                                           &    \centering{08 54 06.0}                                             &        \centering{17.41}            &          \centering{1.23}           &        \centering{20756$\,\pm\,$29}                     &  \multicolumn{1}{r}{3.17}                         &  \multicolumn{1}{r}{ 5.7$\,\pm\,$1.6}                      &     \multicolumn{1}{r}{$-$0.59$\,\pm\,$0.08}              &  \multicolumn{1}{c}{ 42}                  \\
83\dotfill                                              &      \centering{15 49 41.70}                                           &    \centering{08 48 42.1}                                             &        \centering{18.60}            &          \centering{1.11}           &        \centering{19873$\,\pm\,$24}                     &  \multicolumn{1}{r}{8.71}                         &  \multicolumn{1}{r}{2.36$\,\pm\,$0.05}                     &     \multicolumn{1}{r}{$-$0.13$\,\pm\,$0.04}              &  \multicolumn{1}{c}{ 62}                  \\
\hline            
\end{tabular*}     
\begin{footnotesize}
\begin{flushleft}
{\bf Notes:} \\
$^{a}$ Galaxy ID referring to Fig.\ \ref{q3}.\\
$^{b}$ Right ascension and declination shown in $h\,m\,s$.\\
$^{c}$ Model magnitudes from SDSS DR5.\\
$^{d}$ Radial velocities and corresponding confidence parameters $R$ determined with the {\tt IRAF xcsao} package.\\
$^{e}$ SSP ages, metallicites and velocity dispersions determined with the software package  ULySS.\\
\end{flushleft}
\end{footnotesize}
\end{minipage}
\end              {table*}

\end{document}